\documentclass[aps,twocolumn,superscriptaddress,footinbib]{revtex4}  
\usepackage{epstopdf}
\usepackage{graphicx}  
\usepackage{bm}        
\usepackage{braket}
\usepackage{exscale}                  
\usepackage[intlimits]{amsmath}       
\usepackage{amsfonts}
\usepackage{amssymb,amscd}
\usepackage{color}
\usepackage{hyperref}
\usepackage[normalem]{ulem}
\usepackage{slashed}
\usepackage{leftidx}
\usepackage{xcolor,soul}
\usepackage{txfonts}
\usepackage{charter}
\usepackage{dsfont}
\usepackage{mathrsfs}
\usepackage[export]{adjustbox} 
\newtheorem{theorem}{Theorem}
\newtheorem{lemma}{Lemma}

\newtheorem{definition}{Definition}

\usepackage[justification=raggedright,singlelinecheck=false]{caption}

\usepackage{times}

\hypersetup{colorlinks=true, citecolor=blue, urlcolor=blue, linkcolor=blue}
\begin{document}

\title{Quantum Speed Limits Based on the Sharma-Mittal Entropy}

\author{Dong-Ping Xuan}
\affiliation{School of Mathematical Sciences, Capital Normal University, Beijing 100048, China}
\author{Zhi-Xi Wang\footnote{E-mail: wangzhx@cnu.edu.cn}}
\affiliation{School of Mathematical Sciences, Capital Normal University, Beijing 100048, China}
\author{Shao-Ming Fei\footnote{E-mail: feishm@cnu.edu.cn}}
\affiliation{School of Mathematical Sciences, Capital Normal University, Beijing 100048, China}

\begin{abstract}
\textbf{Quantum speed limits (QSLs) establish intrinsic bounds on the minimum time required for the evolution of quantum systems. We present a class of QSLs formulated in terms of the two-parameter Sharma-Mittal entropy (SME), applicable to finite-dimensional systems evolving under general nonunitary dynamics. In the single-qubit case, the QSLs for both quantum channels and non-Hermitian dynamics are analyzed in detail. For many-body systems, we explore the role of SME-based bounds in characterizing the reduced dynamics and apply the results to the XXZ spin chain model. These entropy-based QSLs characterize fundamental limits on quantum evolution speeds and may be employed in contexts including entropic uncertainty relations, quantum metrology, coherent control and quantum sensing.
}
\end{abstract}
\maketitle

\vspace{.4cm}

\section{Introduction}\label{sect1}
The quantum speed limit (QSL) specifies the highest rate that quantum mechanics allows for the evolution of a quantum system subjected to a particular set of dynamics
\cite{MT,ML,PhysRevLett.65.1697,PhysRevLett.120.070401}. It sets the least amount of time that a quantum system needs in order to transition from an initial state to a final state during its evolution. QSLs are essential in advanced quantum technologies, including quantum computing~\cite{NewJPhys.24.065003}, quantum metrology~\cite{QuantumSciTechnol3025002}, optimal control~\cite{PhysRevLett118100601} and quantum thermodynamics~\cite{PhysRevE97062116}, where QSLs play a key role in optimizing performance and enhancing operational efficiency.

The concept of QSL was initially introduced for orthogonal states $|{\psi_0}\rangle$ and $|{\psi_{\tau}}\rangle$ for closed quantum systems through the Mandelstam-Tamm (MT) bound~\cite{MT}, formulated as $T_{\perp} \geq {\hbar\pi}/{\sqrt{{\Delta E_\psi}^2}}$, where the energy variance is given by ${\Delta E_\psi}^2 = \langle H^2 \rangle_\psi - \langle H \rangle_\psi^2$, with $H$ denoting the system Hamiltonian. Subsequently, Margolus and Levitin (ML) derived an alternative expression for the QSL~\cite{ML}, $T_{\perp} \geq {\hbar\pi}/{2 E_\psi}$, where the mean energy is defined as $E_\psi = \langle H \rangle_\psi - E_0$, and $E_0$ is the ground state energy of the system. It is worth noting that the MT and ML bounds differ only in the specific choice of energy reference. Thus, the QSL time for the evolution between two orthogonal pure states,  $|{\psi_0}\rangle$ and $|{\psi_{\tau}}\rangle$, is given by
$$
T_{\text{QSL}} = \max\left\{ \frac{\hbar \pi}{\sqrt{{\Delta E_\psi}^2}}, \frac{\hbar \pi}{2 E_\psi} \right\}.
$$


In the context of closed quantum systems evolving under unitary dynamics generated by
a (time--independent) Hamiltonian, generalized QSLs for transitions between mixed states
$\rho \to \sigma$ were introduced in Ref.~\cite{PhysRevX6021031}. The bound reads
$$
T(\rho \rightarrow \sigma) \geq \frac{\arccos \sqrt{F(\rho, \sigma)}}{\min \left\{ \sqrt{\Delta {E_\rho^2}}, E_\rho\right\}},
$$
where $\sqrt{F(\rho,\sigma)}=\mathrm{tr}\!\left[\sqrt{\sqrt{\rho}\,\sigma\,\sqrt{\rho}}\right]$
is the root of the Uhlmann fidelity between $\rho$ and $\sigma$,
$\Delta E_\rho^{\,2}$ denotes the energy variance with respect to the system Hamiltonian,
and $E_\rho$ is the mean energy above the ground state.

Deffner and Lutz extended the energy--time uncertainty relation to driven closed quantum systems, deriving QSLs for arbitrary initial and final mixed states under arbitrary time--dependent unitary driving~\cite{JMP335302}. Using the Bures angle
$\mathcal{B}(\rho_0,\rho_\tau)=\arccos\!\sqrt{F(\rho_0,\rho_\tau)}$ as a distinguishability measure and bounding its time derivative, they obtained a unified MT and ML-type bound,
\begin{equation}
  \tau_{\mathrm{QSL}}
  = \max\!\left\{
      \frac{\hbar\,\mathcal{B}(\rho_0,\rho_\tau)}{\overline{E}},
      \ \frac{\hbar\,\mathcal{B}(\rho_0,\rho_\tau)}{\overline{\Delta E}}
    \right\},\nonumber
\end{equation}
where $\overline{E}=\tau^{-1}\!\int_{0}^{\tau} E(t)\,dt$ and
$\overline{\Delta E}=\tau^{-1}\!\int_{0}^{\tau} \Delta E(t)\,dt$ are the time-averaged mean energy and energy variance, respectively. The result recovers the standard MT or ML limits for time-independent generators and highlights the strong dependence of the speed limit on the driving protocol.
The QSL reflects the fact that the geodesic $\mathcal{L}(\rho_0, \rho_T)$ between an initial  quantum state $\rho_0$ and a final quantum state $\rho_T$ is the shortest among all physical evolution paths $\ell_\gamma(\rho_0, \rho_T)$, $\mathcal{L}(\rho_0, \rho_T) \leq \ell_\gamma(\rho_0, \rho_T)$~[10]. The QSL bound is saturated when the actual evolution follows the geodesic. QSLs are closely linked to the geometric properties of quantum states, including Fisher information, quantum distances~\cite{PRA110042425,NewJPhys24065001} and the Bures angle~\cite{SciRep105500}. They have been widely studied through the analysis of the dynamics of physical quantities, such as expectation values~\cite{PRA106042436}, autocorrelation functions~\cite{Quantum6884}, and quantum features like quantumness~\cite{srep38149}, coherence~\cite{NewJPhys.24.065003,PRA110042425}, entanglement~\cite{PhysRevA.104.032417} and correlations~\cite{PhysRevA.107.052419}. Moreover, QSLs have been investigated in connection with quantum resources~\cite{NewJPhys24065001} and various information-theoretic measures~\cite{NewJPhys.24.065003,PhysRevE.103.032105}.

The geometric framework has revealed a deep connection between QSLs and measures of quantum state distinguishability. Within this context, there is growing interest in QSLs formulated using entropic quantifiers. Examples include speed limits associated with the von Neumann entropy~\cite{NewJPhys.24.065003}, as well as QSLs constructed from Umegaki relative entropy, applicable to both unitary and nonunitary evolutions, and to single and multi-particle systems~\cite{NewJPhys24065001,PhysRevA.110.052420,PhysRevA.107.052419,PhysRevA.104.052223}. Furthermore, a distinct class of QSLs based on unified entropies has been established for arbitrary dynamical processes~\cite{PhysRevA.106.012403}. The relationship between QSLs and Sharma-Mittal entropy (SME) remains largely unexplored~\cite{SharmaMittalJMathSci1975,S.Mazumdar2019}. As a two-parameter family, SME unifies several prominent entropic measures, including Shannon~\cite{von Neumann entropy}, R\'{e}nyi~\cite{renyi1961measures} and Tsallis entropies~\cite{tsallis1988possible}.

The Sharma--Mittal entropy (SME), a two--parameter generalized entropy, offers distinct advantages for unifying multiple entropic frameworks~\cite{SharmaMittalJMathSci1975,S.Mazumdar2019}. For a density operator $\rho$ with eigenvalues $\{\lambda_i\}$, the SME is
$
\text{S}_{q,z}(\rho)
= \frac{1}{\,z-1\,}\!\left[\!\left(\sum_{i} \lambda_i^{\,q}\right)^{\!\frac{1-z}{\,1-q\,}} - 1\right].
$
The parameter $q$ adjusts the emphasis on high--probability versus low--probability  outcomes, whereas $z$ controls the outer aggregation behavior (and thus the curvature or nonadditivity) of the measure. The SME continuously recovers the R\'enyi entropy as $z \to 1$, the Tsallis entropy as $z \to q$, and reduces to the von Neumann (Shannon) case as $q,z \to 1$. As a purely spectral quantity, $\text{S}_{q,z}$  depends only on state eigenvalues. Although SME has been applied in thermodynamics and information theory and used to quantify quantum correlations such as Sharma--Mittal discord~\cite{SMEdiscord2024}, its role in gauging quantum evolution speed is still underexplored. Building on its spectral nature, we develop SME--based QSLs in terms of spectra and Schatten speeds for general nonunitary dynamics and present the first such bounds while revealing a trade-off between entropy production and evolution speed in many--body systems.


In this paper, we investigate a class of generalized entropic QSLs based on SME for quantum systems undergoing nonunitary dynamics. Our bounds depend only on the eigenvalues of the system's states and the Schatten speed. The findings are implemented in the context of quantum channels as well as non-Hermitian quantum systems, with single-qubit simulations to illustrate the results. Then we demonstrate that the QSL associated with marginal states is governed by the SME and the variance of the Hamiltonian for many-body systems. These results establish entropy-based QSLs applicable to general nonunitary dynamics.


This paper is structured as follows. Section~\ref{sect1} reviews the SME and basic aspects of quantum dynamics. Section~\ref{sect2} derives an upper bound for the SME. Section~\ref{sect3} presents entropy--based QSLs for nonunitary evolutions, with applications to quantum channels and non-Hermitian systems. Section~\ref{sect4} extends the results to many-body systems, illustrated by the XXZ model. Conclusions are drawn in Section~\ref{sect5}.

\section{Preliminaries}\label{sect1}
This section addresses a background on Sharma-Mittal entropy and a brief overview of the quantum dynamical evolution.

\subsection{Sharma-Mittal Entropy}\label{secSME}
Let $\mathcal{H}$ be a Hilbert space with dimension $d = \dim \mathcal{H}$. The collection of quantum states $\mathcal{D}$ comprises all $d \times d$ Hermitian, positive semidefinite matrices with unit trace,
\[
 \mathcal{D}= \{ \rho \in \mathcal{H} \mid \rho^\dagger = \rho,~ \rho \geq 0,~ \text{tr}(\rho) = 1 \}.
\]
Let $\rho = \sum_{i=1}^{d} \lambda_i \lvert \lambda_i \rangle \langle \lambda_i \rvert$ be the eigen-decomposition of $\rho$, with $\lambda_i$ and $\lvert \lambda_i \rangle$ the eigenvalues and eigenvectors, respectively, $0 \leq \lambda_i \leq 1$ and $\sum_{i=1}^{d} \lambda_i = 1$.

\begin{definition}[~\cite{SharmaMittalJMathSci1975,S.Mazumdar2019}]
\label{def:SME}
Let $\rho$ be a density matrix, and $q$ and $z$ two real parameters satisfying $q \in (0,1) \cup (1,+\infty)$ and $z \in (-\infty,1) \cup (1,+\infty)$. The SME of $\rho$ is defined as
\begin{equation}
\text{S}_{q,z}(\rho) = \frac{1}{ z-1} \left[ \left( \sum_{i}^{d} (\lambda_i)^q \right)^{\frac{1 - z}{1 - q}} - 1 \right].
\label{SM1}
\end{equation}
\end{definition}

The $q$-purity is defined by
\[
h_q(\rho) := \mathrm{tr}(\rho^q) = \sum_{i=1}^{d} \lambda_i^q \geq 0,
\]
which is real-valued and non-negative for all $q$.
The SME in \eqref{SM1} can be equivalently rewritten as
\begin{equation}
\text{S}_{q,z}(\rho) = \frac{1}{z-1}
\left[ \left( h_q(\rho) \right)^{\frac{1 - z}{1 - q}} - 1 \right],
\label{SM2}
\end{equation}
where $q \in (0,1) \cup (1,+\infty)$ and $z \in (-\infty,0) \cup (0,+\infty)$.

The SME fulfills the following fundamental properties:
(i) SME is nonnegative, $\text{S}_{q,z}(\rho) \geq 0$;
(ii) SME remains invariant under unitary transformations of the input state, $\text{S}_{q,z}({V}\rho{V^{\dagger}}) = \text{S}_{q,z}(\rho)$ for any unitary $V$ such that $V{V^{\dagger}} = {V^{\dagger}}V = \mathbb{I}$, for values of $q$ within the range $(0,1)\cup(1,+\infty)$ and values of $z$ in the range $(-\infty,1)\cup(1,+\infty)$;
(iii) The SME adheres to the upper bound condition $\text{S}_{q,z}(\rho) \leq \frac{1}{1 - z} \left[ \left( \operatorname{rank}(\rho) \right)^{1 - z} - 1 \right]$ for all $q\in(0,1)$ and $z \in (-\infty,1)\cup(1,+\infty)$. This bound follows from the property that, for any density matrix $\rho$ with finite rank and $q\in(0,1)$, the inequality $\operatorname{tr}(\rho^q) \leq [\operatorname{rank}(\rho)]^{1 - q}$ holds, with equality attained when
$\rho$ is a maximally mixed state~\cite{bookNielsen}.
Notably, in the limiting case $q \to 1$ and $z \to 1$, this bound recovers to the well-known inequality for the von Neumann entropy, $S(\rho) \leq \ln [\operatorname{rank}(\rho)],$ where $S(\rho) = -\text{tr}(\rho \ln \rho)$~\cite{von Neumann entropy}.

The following limiting behaviors illustrate how the Sharma--Mittal entropy provides a unified framework that interpolates between various generalized entropic forms widely used in quantum information theory.
\begin{itemize}
    \item[(i)] In the limit \( z \to 1 \), the SME reduces to the well-known R\'{e}nyi entropy~\cite{renyi1961measures} defined by
    \begin{equation}
    \text{R}_{q} (\rho)= \lim_{z \to 1} \text{S}_{q,z}(\rho) = \frac{1}{1 - q} \ln \left( h_{q}(\rho) \right),
    \label{Renyi}
    \end{equation}
    where \( q \geq 0 \) and \( q \neq 1 \).

    \item[(ii)] In the case \( z \to q \), the SME recovers the Tsallis entropy~\cite{tsallis1988possible},
    \begin{equation}
    \text{T}_{q}(\rho) = \lim_{z \to q} \text{S}_{q,z}(\rho) = \frac{1}{1 - q} \left( h_{q}(\rho) - 1 \right)
    \label{Tsallis}
    \end{equation}
for \( q \geq 0 \) and \( q \neq 1 \).

    \item[(iii)] The SME converges to the von Neumann entropy~\cite{von Neumann entropy} in the joint limit \( q \to 1 \) and \( z \to 1 \),
     \begin{equation}
    \text{H}(\rho) = \lim_{q \to 1, z \to 1} \text{S}_{q,z}(\rho) = -\text{tr}(\rho \ln \rho).
    \label{Shannon}
    \end{equation}
\end{itemize}
\subsection{Quantum Dynamics}\label{subsectionQuantumDynamics}
The characterization of quantum dynamical evolutions is essential for establishing QSLs.
The time evolution of the system under a general dynamical process is described by the master equation~\cite{JMP1976,CMP1976,PhysRev121920},
\begin{equation}
\dot{\rho}_t=\frac{d \rho_t}{d t} = \mathcal{L}_t(\rho_t),
\end{equation}
where $\rho_t$ denotes the state of the system at time $t$, and $\mathcal{L}_t$ is the associated (possibly time-dependent) Liouvillian superoperator~\cite{ARivas2012}.
For Markovian CPTP semigroups, $\mathcal{L}_t$ reduces to the well-known GKSL generator~\cite{JMP1976,CMP1976}. We keep the discussion general and do not assume Markovianity unless stated otherwise. This framework is particularly relevant in discussing QSLs and the fundamental constraints on quantum dynamics.
The seminal work of Mandelstam and Tamm~\cite{MT} is based on the observation that, for quantum systems undergoing unitary evolution described by the Schr\"{o}dinger equation, the dynamics of any observable \( A \) obeys the Liouville--von Neumann equation,
\begin{equation}
\frac{\partial A}{\partial t} = \frac{i}{\hbar} [H, A],
\end{equation}
where \( H \) is the Hamiltonian of the system.

\section{Upper Bound of the Sharma--Mittal Entropy}\label{sect2}
Let the quantum system be initially prepared in a state $\rho_0 \in \mathcal{D}$. The system evolves under a time-dependent, nonunitary dynamical map ${{\mathcal{E}}_t}(\cdot)$, $t\in[0,\tau]$. We consider that the resulting state at time $t$, $\rho_t = {\mathcal{E}}_t(\rho_0)$, is full ranked. In what follows, we restrict ourselves to the parameter regime $1 > z \geq q > 0$ and set $\hbar = 1$ for convenience.

The derivative of $\text{S}_{q,z}({\rho_t})$ with respect to time satisfies
\begin{equation}
\left|\frac{d}{dt}{\text{S}_{q,z}}({\rho_t})\right| = \frac{1}{|1 - q|} \, {{\left[{{h}_q}({{\rho}_t})\right]}^{\frac{q-z}{1-q}}} \left|\frac{d}{dt}{{h}_{q}}({\rho_t})\right|. \label{derivativeSME1}
\end{equation}
For $q \in (0,1)$, the $q$-purity satisfies ${h_{q}}(\rho) \geq 1$.
In this parameter regime, the inequality $[{h_{q}}(\rho)]^{\frac{q-z}{1-q}} \leq 1$ holds whenever $0 < q < z < 1$ or $z > 1$, while $[{h_{q}}(\rho)]^{\frac{q-z}{1-q}} \geq 1$ applies when $0 < z < q < 1$. Conversely, for $q \geq 1$, one has ${h_{q}}(\rho) \leq 1$, and it follows that $[{h_{q}}(\rho)]^{\frac{q-z}{1-q}} \geq 1$ whenever $q > z > 1$, whereas $[{h_{q}}(\rho)]^{\frac{q-z}{1-q}} \leq 1$ for $z > q > 1$.

\begin{figure}[!t]
\begin{center}
\includegraphics[scale=1.]{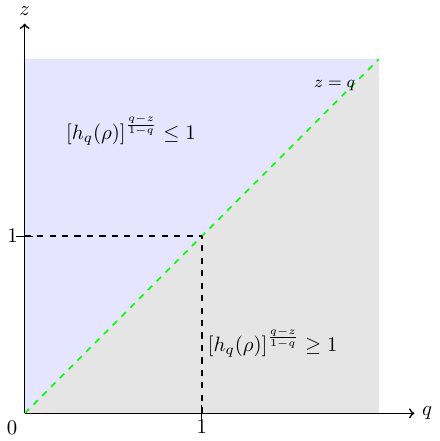}
\end{center}
\caption{
The phase structure of the function $[{h_q}(\rho)]^{\frac{q-z}{1-q}}$ exhibits distinct behaviors depending on the parameters $q$ and $z$.
For $q \in (0,1)$, the $q$-purity satisfies ${h_q}(\rho) \geq 1$.
In this range, the inequality $[{h_q}(\rho)]^{\frac{q-z}{1-q}} \leq 1$ holds for $0 < q < z < 1$ as well as for $z > 1$. When $0 < z < q < 1$, the inequality $[{h_q}(\rho)]^{\frac{q-z}{1-q}} \geq 1$ holds.
In contrast, for $q \geq 1$, the $q$-purity fulfills ${h_q}(\rho) \leq 1$.
Under this condition, the relation $[{h_q}(\rho)]^{\frac{q-z}{1-q}} \geq 1$ holds when $q > z > 1$, while $[{h_q}(\rho)]^{\frac{q-z}{1-q}} \leq 1$ holds for $z > q > 1$.}
\label{functionh_q}
\end{figure}

Fig.~\ref{functionh_q} displays the phase diagram of the function ${[{h_q}(\rho)]^{\frac{q-z}{1-q}}}$ in the $q$-$z$ parameter space.
Notably, for $q \in (0,1)$, the $q$-purity satisfies ${h_q}(\rho) \geq 1$ for any $\rho \in \mathcal{D}$.
Within this regime, the inequality ${[{h_q}(\rho)]^{\frac{q-z}{1-q}}} \leq 1$ holds whenever $1 > z \geq q > 0$~\cite{RasteginJStatPhys2011}.
Substituting this result into \eqref{derivativeSME1}, we immediately obtain
\begin{equation}
\label{derivativeSME2}
\left| \frac{d}{dt}{\text{S}_{q,z}}({\rho_t}) \right| \leq \frac{1}{\left| 1 - q\right| } \left| \frac{d}{dt}{{h}_{q}}(\rho_t) \right|.
\end{equation}
According to Ref.~\cite{PhysRevA.106.012403}, for an arbitrary nonsingular $\rho_t$ and $q\in(0,1)$, the magnitude of the derivative of $h_q({\rho_t})$ is constrained by
\begin{equation}
\left| \frac{d}{dt}{h_q}({\rho_t}) \right| \leq \left({\lambda_{\text{min}}}({\rho_t}) + 1 - q \right) {({ \lambda_{\text{min}}}({\rho_t}))^{q - 2}} \, {\left\|\frac{d{\rho_t}}{dt}\right\|_1}.
\label{eqrhot}
\end{equation}
Here, the Schatten 1-norm defined by ${\|\hat{\mathcal{M}}\|_1} = \mathrm{tr}\left(\sqrt{\mathcal{M}^\dagger \mathcal{M}}\right)$ quantifies the trace norm of an operator $\mathcal{M}$. The term $\lambda_{\mathrm{min}}(\rho_t)$ denotes the minimal eigenvalue of the evolved density matrix $\rho_t$. Notably, Eq.\eqref{eqrhot} features the trace speed ${\left\| \frac{d \rho_t}{dt} \right\|_1}$, often referred to as the Schatten speed \cite{PhysRevA.97.022109}.


The bound in (\ref{eqrhot}) plays a crucial role in the derivation of QSLs, as it links the rate of change of $q$-purity to the dynamical properties of the system, thereby providing an operational handle to characterize the evolution time under nonunitary processes.
Combining \eqref{derivativeSME2} and~\eqref{eqrhot}, we get that the time derivative of the SME is constrained by the following upper bound,
\begin{equation}
\label{derivativeSME3}
\left| \frac{d}{dt}{\text{S}_{q,z}}({\rho_t}) \right| \leq {g_{q}}[{\lambda_{\text{min}}}({\rho_t})]{{\left\| \frac{d{{\rho}_t}}{dt} \right\|}_1} ~,
\end{equation}
where  the auxiliary function
\begin{equation}
\label{g_q}
{g_{q}}[s] := \frac{1}{1 - q}\left(1 - q + s \right) {s^{q - 2}}.
\end{equation}

\begin{theorem}
Consider a general time-dependent nonunitary quantum evolution described by $\rho_t = \mathcal{E}_t(\rho_0)$ with $t \in [0, \tau]$. The Sharma-Mittal entropy of the initial state $\rho_{0}$ and that of the final state $\rho_{\tau}$ obeys the following upper bound:
\begin{equation}
\label{derivativeSME4}
\left| \text{S}_{q,z}(\rho_{\tau}) - \text{S}_{q,z}(\rho_0) \right| \leq \int_0^{\tau} dt\, g_q\big[\lambda_{\text{min}}(\rho_t)\big] \left\| \frac{d \rho_t}{dt} \right\|_1,
\end{equation}
where $g_q(\cdot)$ is the function defined in \eqref{g_q} and $\left\| \mathcal{M} \right\|_1 $ denotes the trace norm.
\label{theorem1}
\end{theorem}

$\mathit{Proof}$. The inequality \eqref{derivativeSME4} follows directly from  \eqref{derivativeSME1}-\eqref{derivativeSME3}. By integrating both sides of \eqref{derivativeSME3} over the time interval $[0, \tau]$ and applying the standard inequality $\left| \int dt\, k(t) \right| \leq \int dt\, |k(t)|$, we obtain (\ref{derivativeSME4}).
$\Box$


The term $\left\| \frac{d \rho_t}{dt} \right\|_1$ on the right side of \eqref{derivativeSME4}  characterizes the quantum speed associated with nonunitary dynamics and can be evaluated once the explicit form of the evolution $\rho_t = \mathcal{E}t(\rho_0)$ is specified.
Moreover, the weight function $g_q[\lambda_{\mathrm{min}}(\rho_t)]$ depends solely on the  $\lambda_{\mathrm{min}}(\rho_t)$ and is parameterized by the entropic parameter $q$.
It should be noted that the left part of \eqref{derivativeSME4} gauges the discriminability between the initial state $\rho_0$ and final state $\rho_{\tau}$  via the absolute difference of their SMEs. This value is upper-bounded by the average rate dictated by $\left\| \frac{d \rho_t}{dt} \right\|_1$ associated with the variation of the instantaneous state.
Notably, evaluating this bound is computationally efficient, as it depends only on the $\lambda_{\mathrm{min}}(\rho_t)$. This feature is particularly beneficial for analyzing SMEs in high-dimensional settings, such as quantum many-body systems for which calculating the entire eigenvalue spectrum of the density matrix usually entails significant computational costs.

To further assess the tightness of the SME bound established in Theorem~\ref{theorem1}, we introduce the following definition.

\begin{definition}
\label{definitionerror}
We define the following normalized relative error to quantify the tightness of the bound on SME,
\begin{equation}
\widetilde{\varsigma}_{q,z}(\tau) := \frac{ \varsigma_{q,z}(\tau) - \min(\varsigma_{q,z}) }{ \max(\varsigma_{q,z}) - \min(\varsigma_{q,z}) },
\label{tightness1}
\end{equation}
where \( \varsigma_{q,z}(\tau) \) denotes the (unnormalized) relative error,
\begin{equation}
\varsigma_{q,z}(\tau):= 1 - \frac{\left| \triangle\text{S}_{q,z}\right|}{\Gamma},
\label{tightness2}
\end{equation}
with
\[
\left|\triangle \text{S}_{q,z}\right| = \left| \text{S}_{q,z}(\rho_{\tau}) - \text{S}_{q,z}(\rho_0) \right|,~~
\Gamma = \int_0^{\tau} dt\, g_{q}\left[\lambda_{\text{min}}(\rho_t)\right] \left\| \frac{d \rho_t}{dt} \right\|_1.
\]
Here, $\Gamma$ is given by the function $g_q$ of the minimal eigenvalue $\lambda_{\text{min}}(\rho_t)$ of the instantaneous state, as well as the trace speed of $\rho_t$.
\end{definition}

Overall, a smaller relative error defined in \eqref{tightness1} implies a tighter bound on the SME given in \eqref{derivativeSME4}. The normalized relative error takes values in $0 \leq \widetilde{\varsigma}_{q,z}(\tau) \leq 1$.
We would like to indicate that the concept of tightness considered here differs from the geometric interpretation discussed in Refs.~\cite{PhysRevLett.65.1697,JMP451787,PhysRevX6021031,PRA110042425,NewJPhys24065001},
where for an evolution from an initial state $\rho_0$ to a final state $\rho_{\tau}$, the QSL bound is deemed saturated when the system's actual evolution path is identical to the geodesic that links these two states. In the present context, the tightness of the bound in \eqref{derivativeSME4} is described by the relative error given in \eqref{tightness1}. This relative error measures the divergence between the absolute change in the SMEs and the integrated quantum speed. As a result, the bound is saturated when the rate of variation of the SME precisely equals to the product of the weight function \(g_q\big[\lambda_{\text{min}}(\rho_t)\big]\) and the instantaneous quantum speed \(\left\|\frac{d\rho_t}{dt}\right\|_1\) generated by the nonunitary dynamics. Subsequently, commencing from \eqref{derivativeSME4}, we deduce a class of QSLs associated with the SME for general nonunitary dynamics.


\section{A Family of Quantum Speed Limits Based on Sharma-Mittal Entropy}\label{sect3}


We study QSLs by using the SME of quantum systems subject to general nonunitary evolutions.

\begin{theorem}
\label{theoremQSL}
The evolution duration $\tau$ admits the following lower bound under a general time-dependent nonunitary evolution from an initial state $\rho_0$ to a final state $\rho_\tau$ within the time interval $t \in [0, \tau]$,
\begin{equation}
\label{QSL1}
\tau \geq \tau^{\text{QSL}}_{q,z}\equiv\frac{\left|\triangle \text{S}_{q,z}\right|}{\langle \Gamma \rangle_\tau}
\end{equation}
with $\langle \Gamma \rangle_{\tau}={\Gamma}/{\tau}$ and \( 0 < q < z < 1 \).
\end{theorem}

$\mathit{Proof.}$
\eqref{QSL1} follows directly by integrating both sides of \eqref{derivativeSME4} over the interval $[0, \tau]$ and then dividing by $\tau$.
$\Box$


The QSL presented in \eqref{QSL1} inherently depends on time, as it characterizes the evolution between two general quantum states $\rho_0$ and $\rho_{\tau}$. Importantly, the absolute difference between their corresponding SMEs serves as a quantification of distinguishability, effectively offering a geometric perspective by interpreting this quantity as a functional distance between the initial and final states.


Furthermore, the QSL is governed by $\left\| \frac{d \rho_t}{dt} \right\|_1$ which encapsulates the rate at which the quantum state evolves under the nonunitary dynamics within the interval $t \in [0, \tau]$. As a result, the quantity $\tau_{\text{QSL}}$ reflects the dynamical behavior of the eigenstates associated with the generators of the evolution~\cite{Deffner2017}. Notably, the QSL time exhibits an inverse relationship with the average evolution speed, weighted by a function that explicitly depends on $\lambda_{\text{min}}(\rho_t)$. It is worth emphasizing that the accuracy of the QSL given in \eqref{QSL1} is inherently related to the strictness of the inequality demonstrated in \eqref{derivativeSME4}. As a consequence, the relative error introduced by \eqref{tightness2} offers a quantitative measure for evaluating how closely the QSL bound in \eqref{QSL1} approximates the actual evolution speed.

Prior QSLs for open dynamics are commonly formulated in terms of state distances and generator norms. For Markovian Lindblad semigroups, the adjoint Lindbladian replaces the Hamiltonian in an MT-type bound, and the extensions to general CPTP maps have been provided~\cite{PhysRevLett.110.050403}.
More recently, a thermodynamic decomposition of the velocity into energy fluctuations, entropy production with dynamical activity, and a bath--induced unitary term has been established~\cite{FunoNJP2019}.
In contrast, our bounds are entropic: by upper--bounding the time derivative of the SME, Theorem~\ref{theoremQSL} leads to a family of QSLs  that depend only on spectral data (including the minimal eigenvalue) and on the Schatten speed, without requiring explicit knowledge of the Lindblad generator or any Kraus--operator optimization. This makes the bounds broadly applicable to general nonunitary evolutions (quantum channels and effective non--Hermitian dynamics are treated explicitly), while our many--body analysis further reveals an inverse dependence on the Hamiltonian variance, consistent with MT-type intuition.
Accordingly, the present SME-based perspective complements distance-based and thermodynamic QSLs by offering a tunable, spectrum--only framework that can tighten bounds across different noise strengths and spectral profiles.


We now examine the limiting behavior of the QSL expression in \eqref{QSL1}, focusing on several representative and physically relevant parameter regimes.
When \( z \to 1 \), the SME reduces to the R\'{e}nyi entropy~\cite{renyi1961measures}, leading to a QSL time
\[
\tau_{q, z \to 1}^{\text{QSL}} = \frac{\left| \text{R}_{q}(\rho_{\tau}) - \text{R}_{q}(\rho_0) \right|}{\langle \Gamma \rangle_{\tau}},
\]
where \( \text{R}_{q}(\rho) \) denotes the R\'{e}nyi entropy. In the limit \( q \to 1 \), the R\'{e}nyi entropy approaches the von Neumann entropy. Consequently, the weight function diverges, i.e., \( \lim_{q \to 1} g_q(\lambda_{\text{min}}(\rho_t)) \to \infty \) [see \eqref{g_q}], which leads to a vanishing QSL time, \( \lim_{q \to 1} \tau^{\text{QSL}}_{q,1} \approx 0 \).

Similarly, for \( z \to q \), the SME collapses into the Tsallis entropy~\cite{tsallis1988possible}, yielding the QSL time
\[
\tau_{q, z \to q}^{\text{QSL}} = \frac{\left| \text{T}_{q}(\rho_{\tau}) - \text{T}_{q}(\rho_0) \right|}{\langle \Gamma \rangle_{\tau}}.
\]
Analogously, in the limit \( q \to 1 \), the Tsallis entropy converges to the von Neumann entropy, yielding
\( \lim_{q \to 1} \tau^{\text{QSL}}_{q,q} \approx 0 \) as well.
Thus, in both limiting cases, the QSL becomes trivial, i.e., \( \tau \geq 0 \), indicating that the bounds lose sensitivity to the nonunitary dynamics of the system when \( q \to 1 \).

\begin{figure}[!t]
\begin{center}
\includegraphics[scale=1.]{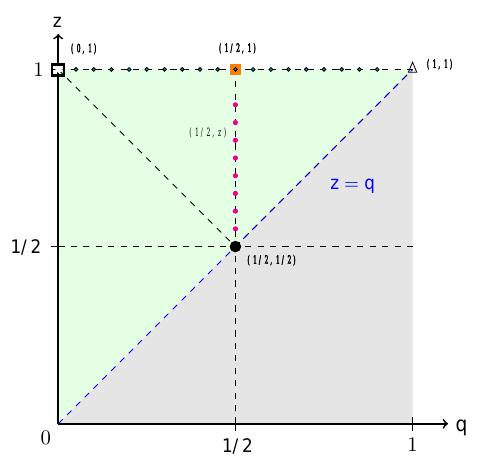}
\end{center}
\caption{Phase diagram for the  quantum speed limit $\tau_{q,z}^{\text{QSL}}$ given in \eqref{QSL1}. The green shaded area indicates the region ($0 < q < z < 1$) in which the QSL applies.}
\label{qzQSL}
\end{figure}

Fig.~\ref{qzQSL} schematically depicts the regions where the QSLs apply. Specifically:
\begin{itemize}
    \item For \( z \to 1 \), the QSL time \( \tau_{q, z \to 1}^{\text{QSL}} \) (blue diamond dotted line) relates to the R\'{e}nyi entropy.
    \item For \( z \to q \), the QSL time \( \tau_{q, z \to q}^{\text{QSL}} \) (blue dotted line) corresponds to the Tsallis entropy.
    \item For \( q = 1/2 \) and \( z \geq 1/2 \), the QSL time \( \tau_{1/2,z}^{\text{QSL}} \) is associated with the SME (pink dotted line),
    \[
    \text{S}_{1/2,z}(\rho) = \frac{1}{1-z} \left[ \left( h_q(\rho) \right)^{2(1-z)} - 1 \right].
    \]
    \item For \( q = 1/2 \) and \( z = 1/2 \), the QSL time \( \tau_{1/2,1/2}^{\text{QSL}} \) relates to
   $\text{S}_{1/2,1/2}(\rho) = 2 \left( h_q(\rho) - 1 \right),$
    which is connected to the $q$-purity (solid black circle).
    \item For \( q = 1/2 \) and \( z = 1 \), the QSL time \( \tau_{1/2,1}^{\text{QSL}} \) corresponds to
    $\text{S}_{1/2,1}(\rho) = 2 \ln \left( h_q(\rho) \right)~~ \text{(orange square mark)}.$
    \item In the special case \( q = z = 1 \), the QSL time \( \tau_{1,1}^{\text{QSL}} \) depends on the von Neumann entropy (black dot mark).
\end{itemize}
Finally, we emphasize that the bounds are not valid for \( q = 0 \) (white rectangular) and \( q = 1, z = 1 \) (white triangle), where the SME is not well defined at these points.


We illustrate the applicability of the results derived in Eqs.\eqref{derivativeSME4}, \eqref{tightness1} and~\eqref{QSL1} in next section, by analyzing two representative nonunitary quantum dynamics, the evolution governed by quantum channels, and the nonunitary dynamics in dissipative quantum systems described by effective non-Hermitian Hamiltonians.

\subsection{QSLs under CPTP Dynamics}\label{channels5}
We consider general open--system evolutions described by CPTP maps. By Stinespring dilation, any such evolution admits a time--dependent Kraus (operator--sum) form
\begin{equation}
  \mathcal{E}_t(\rho)=\sum_{\ell} V_\ell(t)\,\rho\,V_\ell^\dagger(t),\qquad
  \sum_{\ell} V_\ell^\dagger(t)V_\ell(t)=\mathbb{I},\nonumber
\end{equation}
so that the state at time $t$ is $\rho_t=\mathcal{E}_t(\rho_0)$~\cite{Stinespring1955,Choi1975,bookNielsen}. Microscopically, if $U_t=e^{-iHt/\hbar}$ is the unitary on system+environment with initial $\rho_{SE}(0)=\rho_0\otimes\rho_E$, then
$\rho_t=\mathrm{Tr}_E\!\left[U_t(\rho_0\otimes\rho_E)U_t^\dagger\right]
=\sum_{\ell}V_\ell(t)\rho_0 V_\ell^\dagger(t)$.
Conversely, a (time-local) GKLS/Lindblad master equation $\dot\rho_t=\mathcal{L}_t(\rho_t)$ generates a CPTP propagator
$\mathcal{E}_t=\mathcal{T}\exp\!\big(\int_0^t\mathcal{L}_s ds\big)$, which also admits a Kraus decomposition via its Choi matrix~\cite{JMP1976,CMP1976}.
We use the Kraus form because it gives $\rho_t$ explicitly, enabling direct evaluation of
spectral entropies and Schatten speeds for our QSL bounds.

The Schatten speed is upper bounded as follows,
\begin{align}
\left\| \frac{d\rho_t}{dt} \right\|_1 &=\left\| \sum_{l} \frac{d}{dt} \left( V_l\, \rho_0 V_l^\dagger \right) \right\|_1 \notag \\
&\leq \sum_{l} \left\| \frac{d}{dt} \left( V_l\, \rho_0 V_l^\dagger \right) \right\|_1 \notag \\
&= \sum_{l} \left\| \left( \frac{dV_l}{dt} \right) \rho_0 V_l^\dagger + V_l\, \rho_0 \left( \frac{dV_l^\dagger}{dt} \right) \right\|_1 \notag \\
&\leq \sum_{l} \left( \left\| \left( \frac{dV_l}{dt} \right) \rho_0 V_l^\dagger \right\|_1 + \left\| V_l\, \rho_0 \left( \frac{dV_l^\dagger}{dt} \right) \right\|_1 \right) \notag \\
&=  2 \sum_{l}{\|{V_{l}}\,{\rho_0}{\dot{V}_{l}^{\dagger}}\|}_1,
\label{channel1}
\end{align}
where the first inequality follows from the triangle inequality for the trace norm and the second inequality follows from the property ${\| \mathcal{X}^{\dagger} \|}_1 = {\| \mathcal{X} \|}_1$ for any operator $\mathcal{X}$.

Substituting \eqref{channel1} into the general bound in \eqref{derivativeSME4}, we obtain
\begin{equation}
\label{channel2}
\left| \text{S}_{q,z}(\rho_{\tau}) - \text{S}_{q,z}(\rho_0) \right|
\leq 2 \tau \sum_{l} \langle \Gamma' \rangle_{\tau},
\end{equation}
where
\[
\Gamma' := g_q\left[\lambda_{\text{min}}(\rho_t)\right] \left\| V_l \rho_0{\dot{V}_{l}^{\dagger}} \right\|_1.
\]
Based on \eqref{channel2}, the relative error can be expressed as
\begin{equation}
\label{channel3}
\varsigma_{q,z}(\tau) = 1 - \frac{\left| \text{S}_{q,z}(\rho_{\tau}) - \text{S}_{q,z}(\rho_0) \right|}{2 \sum_{l} \int_0^{\tau} dt\, g_q\left[\lambda_{\text{min}}(\rho_t)\right] \left\| V_l \rho_0{\dot{V}_{l}^{\dagger}}
\right\|_1}.
\end{equation}

\begin{theorem}
\label{thm:qsl_cptp}

Under a general time-dependent CPTP map, $\mathcal{E}_t(\cdot) = \sum_{l} V_l \cdot V_l^\dagger$, the evolution time $\tau$ obeys the following lower bound,
\begin{equation}
\tau \geq \tau_{q,z}^{\mathrm{QSL}}=\frac{\left| {{\text{S}}_{q,z}}({\rho_{\tau}}) - {{\text{S}}_{q,z}}({\rho_0})  \right| }{2  \sum_{l} \langle \Gamma'\rangle_{\tau}},
\label{channel4}
\end{equation}
where $\langle \Gamma' \rangle_{\tau}=\int_0^{\tau} \Gamma' dt\,/\tau$ and \( 0 < q < z < 1 \).
\end{theorem}

$\mathit{Proof.}$
Inserting the result from \eqref{channel1} into the general expression in \eqref{QSL1} yields the QSL time formulated in \eqref{channel4}.
$\Box$


Theorem~\ref{thm:qsl_cptp} provides a fundamental lower bound on the evolution time within the scope of a CPTP dynamical mapping. This bound is formulated in terms of the absolute change in the SME, and explicitly incorporates both the time-dependent Kraus operators $\{V_l\}$ that govern the dynamics and the $\lambda_{\text{min}}(\rho_t)$. In the subsequent analysis, we focus on evaluating the relative error defined in \eqref{channel3} along with the QSL bound  given by \eqref{channel4} for single-qubit systems.

We consider a two-level quantum system undergoing  the amplitude damping channel with Kraus operators \cite{PhysRevA.102.012401},
\[
V_0 = |0\rangle\langle{0}| + e^{-\gamma t/2}\, |1\rangle\langle{1}|,~~V_1 = \sqrt{1 - e^{-\gamma t}}\, |0\rangle\langle{1}|.
\]
The system is initially set up in a general single-qubit state,
\begin{align}
\rho_0 = \frac{1}{2}(\mathbb{I} + \mathbf{r} \cdot \boldsymbol{\sigma}),
\label{ini_state}
\end{align}
where the Bloch vector $\mathbf{r}=\{r\sin\theta\cos\phi,r\sin\theta\sin\phi,r\cos\theta\}$ with $r\in [0,1]$, $\theta\in [0,\pi]$ and $\phi\in[0,2\pi)$, and $\boldsymbol{\sigma}=\{\sigma_x,\sigma_y,\sigma_z\}$ is given by the standard Pauli matrices.
The density matrix evolves to
\[
\rho_t = \sum_{l=0,1} V_l \,\rho_0 V_l^{\dagger}
\]
at time $t$.

To compute the QSL bound given by \eqref{channel4} and the relative error defined in \eqref{channel3}, we first calculate the quantity $h_q(\rho_t)$,
\[
h_q(\rho_t) = \left( \lambda_{\text{min}}(\rho_t) \right)^q + \left( 1 - \lambda_{\text{min}}(\rho_t) \right)^q,
\]
where $\lambda_{\text{min}}(\rho_t)$ represents the smallest eigenvalue of $\rho_t$,
\[
\lambda_{\text{min}}(\rho_t) = \frac{1}{2} \left( 1 - \sqrt{1 - e^{-\gamma t} \, \xi_t} \, \right)
\]
with $\xi_t = 1 - r^2 + (1 - e^{-\gamma t})(1 - r \cos\theta)^2$.
In particular, at the initial time $t=0$, the $q$-purity simplifies to
\[
h_q(\rho_0) = 2^{-q} \left[ (1 - r)^q + (1 + r)^q \right].
\]
The absolute value of the difference in the SME is given by
\begin{align*}
&\left|{{\text{S}}_{q,z}}({\rho_{\tau}}) - {{\text{S}}_{q,z}}({\rho_0})\right|=
\frac{1}{1-z}\\
&\left| {\left( {\lambda^{q}_{\text{min}}}({\rho_{\tau}}) + {\left( 1 - {\lambda_{\text{min}}}({\rho_{\tau}})\right)^{q}}\right)^{\frac{1-z}{1-q}}}  - {2^{\frac{q(z-1)}{1-q}}}\left({(1 - r)^{q}} + {(1 + r)^{q}}\right)^{\frac{1-z}{1-q}} \right|,
\end{align*}
and the Schatten speed expressed is given by
\begin{align*}
&\sum_{l}{\|{V_{l}}{\rho_0}{\dot{V}_{l}^{\dagger}}\|}_1=\\
&\frac{1}{4}\gamma  e^{-\gamma t}\left(1 - r\cos\theta + \sqrt{{(1 - r\cos\theta)^2} + {e^{\gamma t}}{r^2}{\sin^2}\theta}\, \right).
\end{align*}
From Theorem~\ref{thm:qsl_cptp}, the corresponding QSL time is
\begin{widetext}
\begin{equation}
\label{channel5}
{\tau^{\text{QSL}}_{q,z}} = \frac{\left| {\left( {\lambda^{q}_{\text{min}}}({\rho_{\tau}}) + {\left( 1 - {\lambda_{\text{min}}}({\rho_{\tau}})\right)^{q}}\right)^{\frac{1-z}{1-q}}}  - {2^{\frac{q(z-1)}{1-q}}}\left({(1 - r)^{q}} + {(1 + r)^{q}}\right)^{\frac{1-z}{1-q}} \right|}{\frac{\gamma}{2(1-q)}\frac{1}{\tau}{\int_0^{\tau}} \, dt \, {e^{-\gamma t}}(1 - q + {\lambda_{\text{min}}}({\rho_t})) \, {{\lambda^{q - 2}_{\text{min}}}({\rho_t})}\left(1 - r\cos\theta + \sqrt{{(1 - r\cos\theta)^2} + {e^{\gamma t}}{r^2}{\sin^2}\theta}\, \right) } ~.
\end{equation}
\end{widetext}

\begin{figure*}[t]
\centering
\newcommand{\panel}[2]{%
  \begin{minipage}[t]{0.32\textwidth}\centering
    \includegraphics[width=\linewidth]{#1}\\[-0.3ex]
    {\small (#2)}
  \end{minipage}%
}
\makebox[\textwidth][c]{%
  \panel{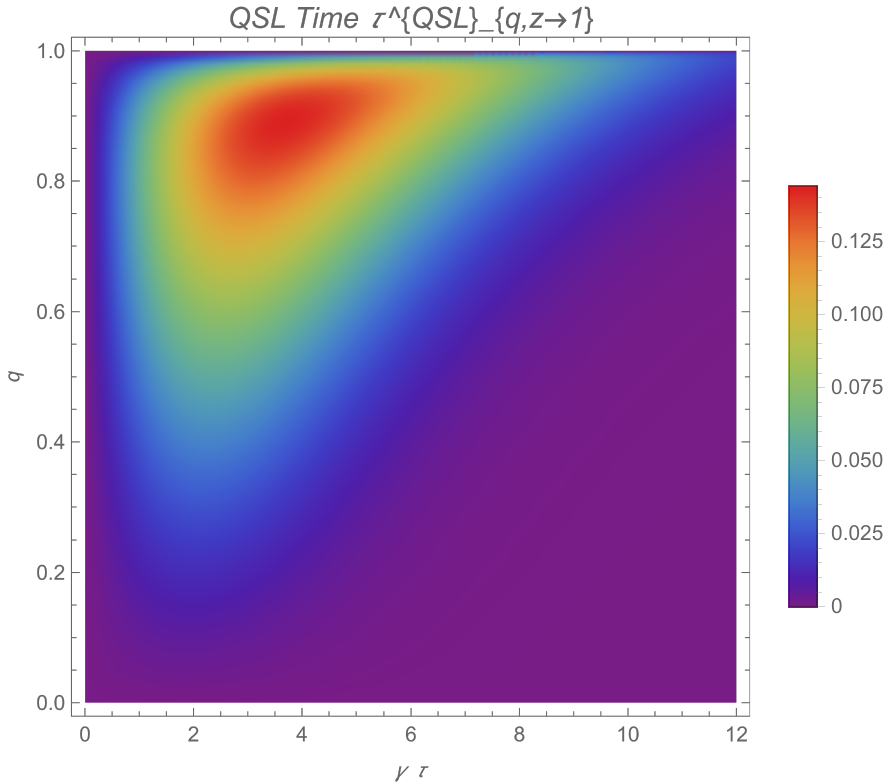}{a}\hfill
  \panel{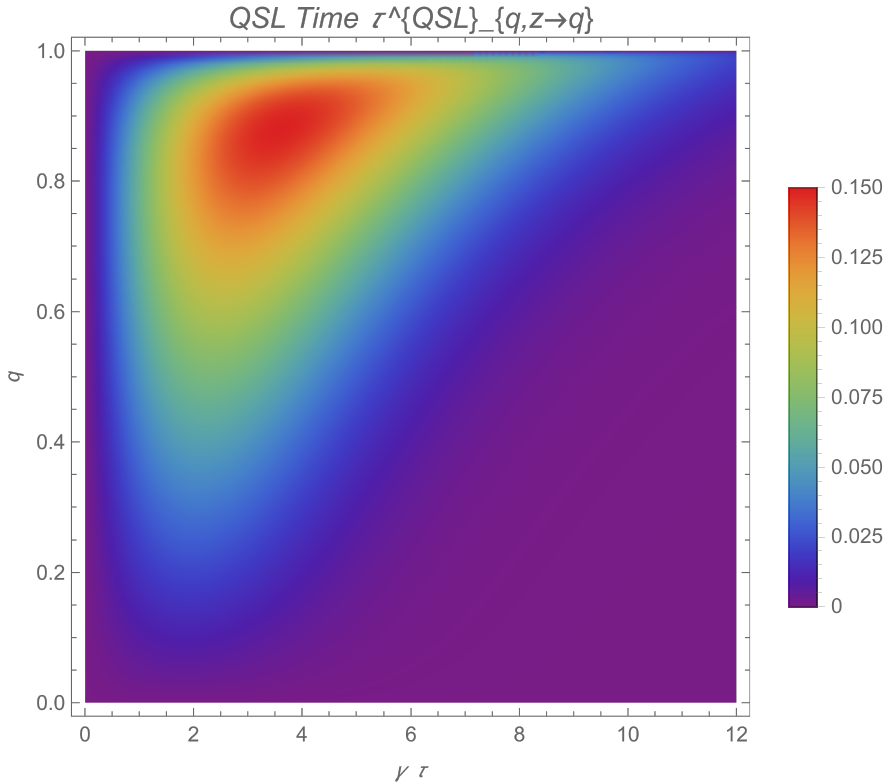}{b}\hfill
  \panel{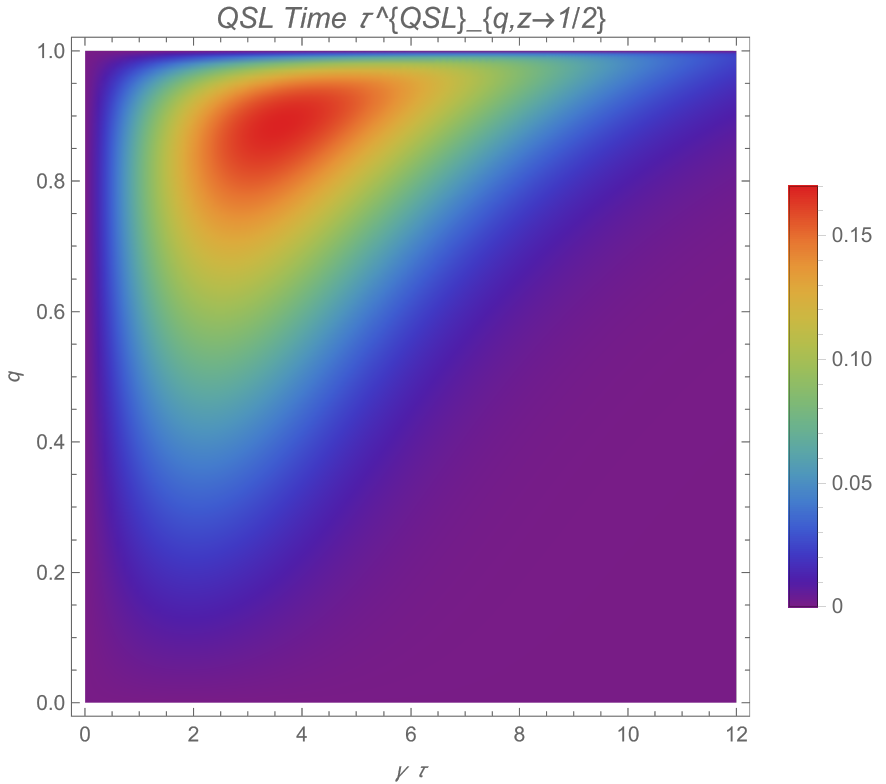}{c}%
}
\vspace{0.8ex}
\makebox[\textwidth][c]{%
  \panel{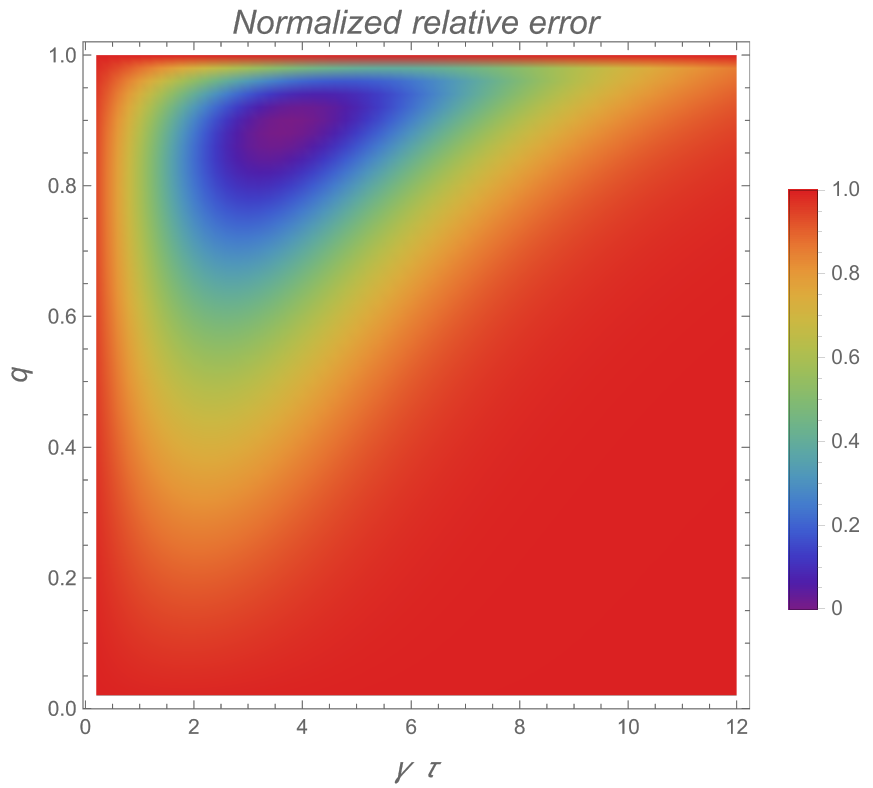}{d}\hfill
  \panel{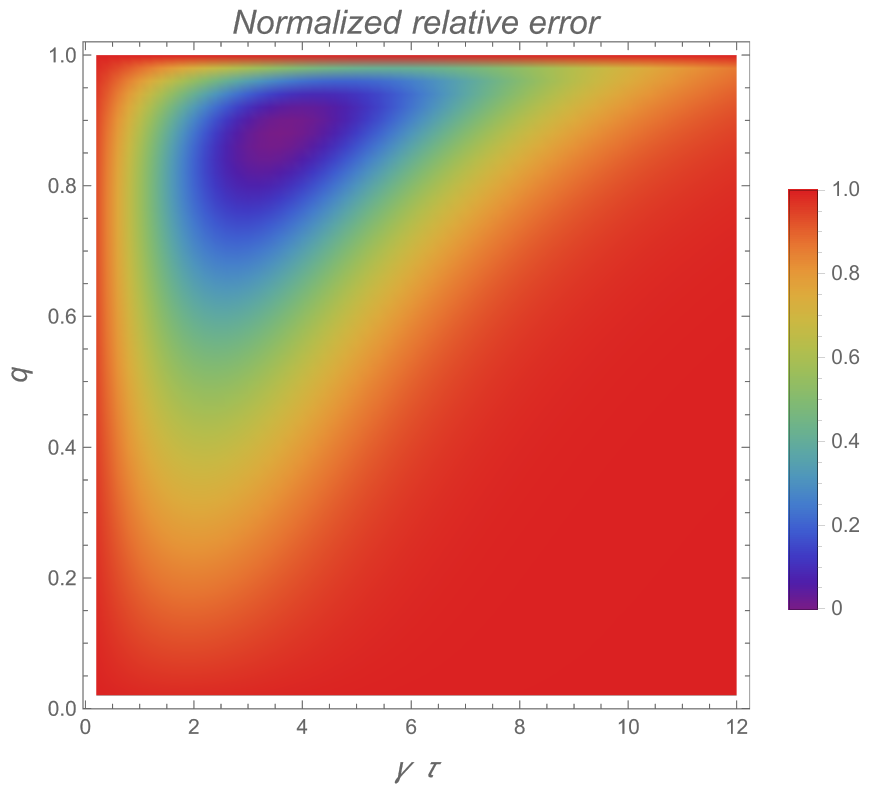}{e}\hfill
  \panel{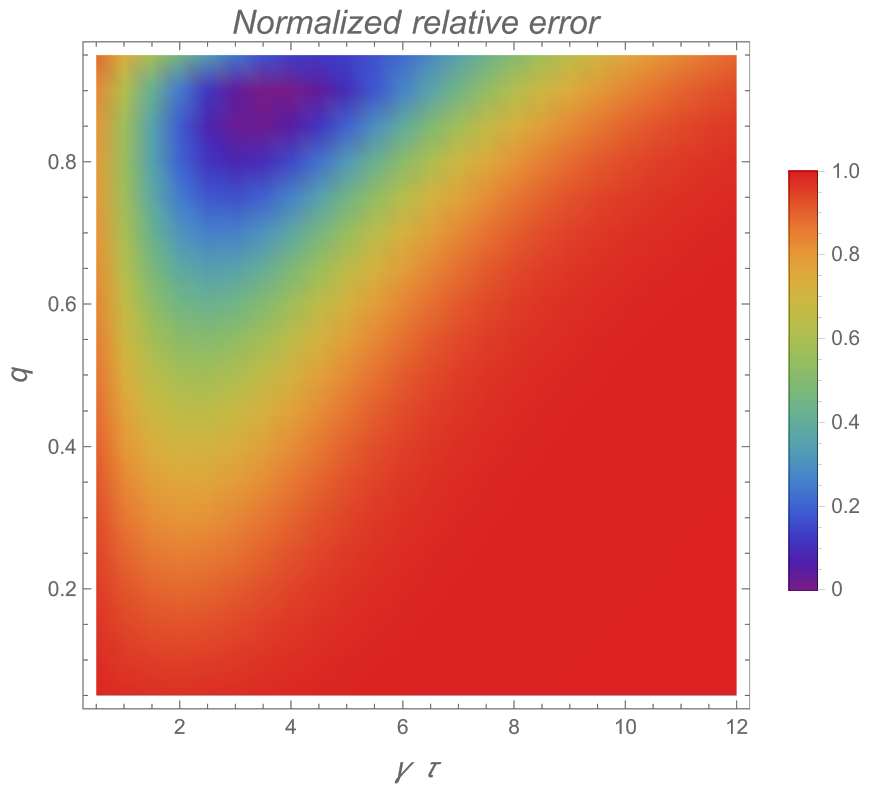}{f}%
}
\caption{%
Density plots illustrate the behavior of the QSL time $\tau^{\text{QSL}}_{q,z}$ [as defined in \eqref{channel5}] and the normalized relative error ${\widetilde{\varsigma}_{q,z}}(\tau)$ [specified by Eq.~\eqref{channel3}] as functions of $\gamma\tau$. The initial state is given by \eqref{ini_state} with Bloch vector $\{r,\theta,\phi\}=\{\tfrac12,\tfrac{\pi}{4},\tfrac{\pi}{4}\}$.}
\label{figchannel1}
\end{figure*}

Fig.~\ref{figchannel1} illustrates the behavior of the QSL time ${\tau^{\text{QSL}}_{q,z}}$ and the normalized relative error $\widetilde{\varsigma}_{q,z}(\tau)$. The system is initially set in a state specified by the parameters $\{r,\theta,\phi\} = \{1/2,\pi/4,\pi/4\}$.
The panels~\ref{figchannel1}(a)-\ref{figchannel1}(c) display the QSL times obtained from \eqref{channel5} corresponding to different entropy measures: the R\'{e}nyi entropy, $\text{R}_{q}(\rho) = \text{S}_{q,z \to 1}(\rho)$; the Tsallis entropy, $\text{T}_{q}(\rho)={\text{S}}_{q, z \to q}(\rho)$; and the Sharma--Mittal entropy with $z = 1/2$, $\text{S}_{q,z = 1/2}(\rho)$. We observe that, for any $q \in (0,1)$, the QSL times $\tau^{\text{QSL}}_{q,z \to 1}$, $\tau^{\text{QSL}}_{q,z \to q}$ and $\tau^{\text{QSL}}_{q,z = 1/2}$ exhibit nonzero values at the earlier stages of the evolution, and gradually decay towards zero as time progresses. Although slight quantitative differences are noted among the three cases, they share the same qualitative behavior. Particularly, the QSL times reach their maximum values within the region $0.6 \lesssim q \lesssim 0.9$ and $2 \lesssim \gamma\tau \lesssim 6$.


In a similar manner, Figs.~\ref{figchannel1}(d)-\ref{figchannel1}(f) display the normalized relative error ${\widetilde{\varsigma}_{q,z}}(\tau)$ corresponding to the R\'{e}nyi, Tsallis and Sharma-Mittal entropies with $z = 1/2$, as specified by \eqref{channel3}. It is found that, across the entire interval $0 < q < 1$, the normalized relative error significantly decreases within $2 \lesssim \gamma\tau \lesssim 6$, and approaches unity at longer times. Specifically, the limit ${\lim_{\tau \to \infty}}\, {\widetilde{\varsigma}_{q,z}}(\tau) \approx 1$ holds for all $q \in (0,1)$, accompanied by $\tau^{\text{QSL}}_{q,z} \approx 0$ as indicated in Figs.~\ref{figchannel1}(a)-\ref{figchannel1}(c). Consequently, this indicates that as $\tau$ becomes large, the lower bound on the actual evolution time becomes increasingly loose, effectively yielding the trivial condition $\tau \gtrsim 0$.

\begin{figure*}[t]
  \centering
  \begin{minipage}[t]{0.32\textwidth}\centering
    \includegraphics[width=\linewidth]{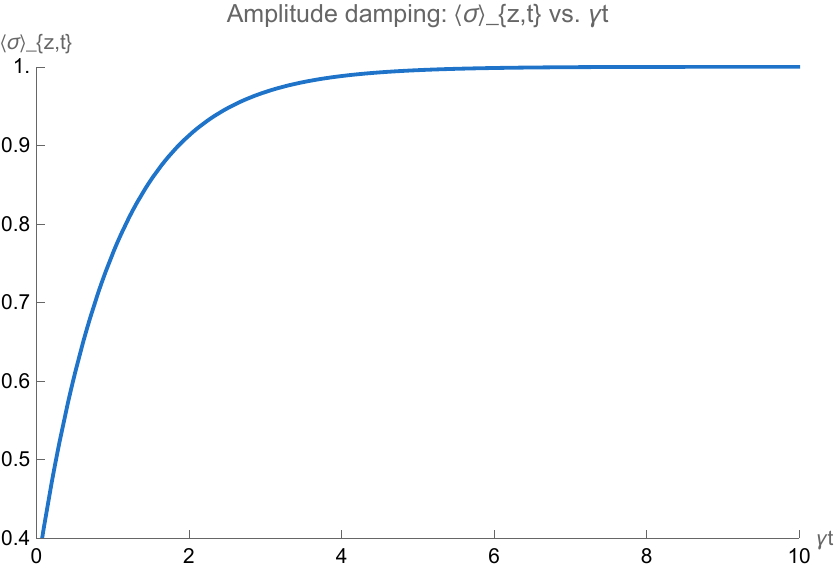}
    \vspace{0.2ex}\\{\small (a) }
  \end{minipage}\hfill
  \begin{minipage}[t]{0.32\textwidth}\centering
    \includegraphics[width=\linewidth]{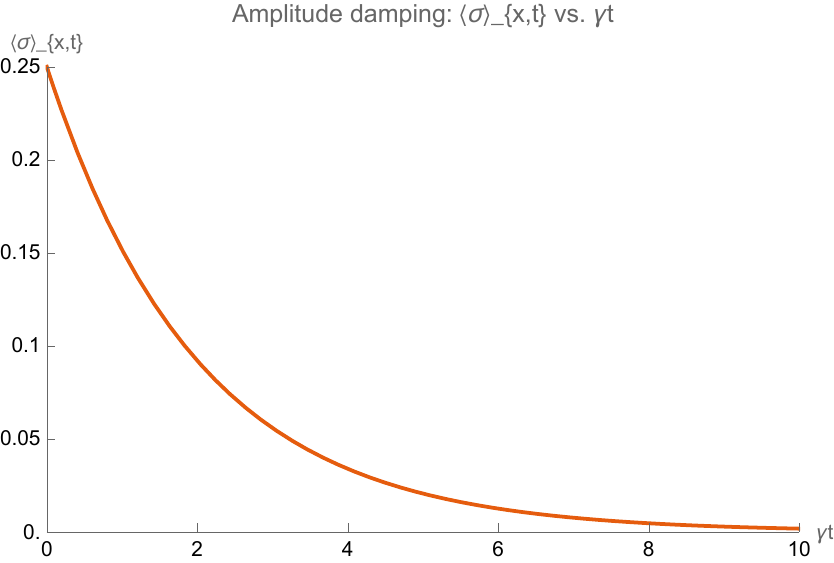}
    \vspace{0.2ex}\\{\small (b)}
  \end{minipage}\hfill
  \begin{minipage}[t]{0.32\textwidth}\centering
    \includegraphics[width=\linewidth]{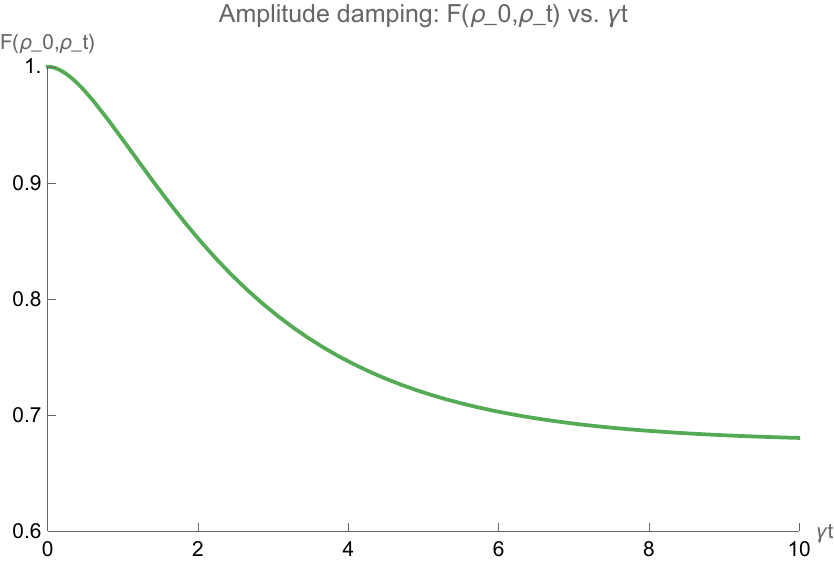}
    \vspace{0.2ex}\\{\small (c) }
  \end{minipage}
  \caption{%
  Physical signatures for the amplitude-damping dynamics (same parameters as Fig.~\ref{figchannel1}).
  (a) The longitudinal polarization $\langle\sigma_z\rangle_t$ tends monotonically to $+1$,
  (b) the transverse polarization $\langle\sigma_x\rangle_t$ decays to $0$,
  and (c) the fidelity $F(\rho_0,\rho_t)$ stays strictly positive for all finite times.}
  \label{fig:AD_diagnostics}
\end{figure*}

$\textit{Physical signatures for the amplitude damping dynamics.}$ To substantiate and complement Fig.~\ref{figchannel1}, we  focus on the single--qubit polarizations and the Uhlmann fidelity along the evolution (scaled time $s=\gamma t$ and same initial state and parameters as in Fig.~\ref{figchannel1} for $\{r,\theta,\phi\} = \{1/2,\pi/4,\pi/4\}$). The Bloch vector evolves as
\[
\mathbf r_t=\big(e^{-s/2}r\sin\theta\cos\phi,\; e^{-s/2}r\sin\theta\sin\phi,\; e^{-s}r\cos\theta+1-e^{-s}\big),
\]
so that $\langle\sigma_x\rangle_t=r_x(t) \to 0$ while $\langle\sigma_z\rangle_t=r_z(t) \to 1$, signaling relaxation to the fixed point $\lvert0\rangle\!\langle0\rvert$. Consistently, the fidelity
\[
F(\rho_0,\rho_t)=\tfrac12\!\left(1+\mathbf r_0\cdot \mathbf r_t+\sqrt{1-\|\mathbf r_0\|^2}\,\sqrt{1-\|\mathbf r_t\|^2}\right)
\]
remains strictly positive at any finite time and approaches $\tfrac12(1+r\cos\theta)=\langle0|\rho_0|0\rangle$ as $t\to\infty$, i.e., no finite--time orthogonalization occurs. Fig.~\ref{fig:AD_diagnostics} shows that, for early and mid times ($\gamma t\!\lesssim\!5$), $\langle\sigma_z\rangle_t$ rises while $\langle\sigma_x\rangle_t$ decays, and the fidelity $F(\rho_0,\rho_t)$ decreases but never vanishes, signaling appreciable state change without orthogonalization. This coincides with Fig.~\ref{figchannel1}, where $\tau^{\mathrm{QSL}}_{q,z}$ is sizable and the normalized relative error is small (tight bound). As the polarizations saturate and the fidelity plateaus at long times, the entropy change becomes negligible, and the SME--based QSL in Fig.~3 correspondingly loosens ($\tau^{\mathrm{QSL}}_{q,z}\!\to\!0$, error $\to 1$).

This phenomenon can be ascribed to the circumstance that the amplitude damping channel gradually squeezes the Bloch sphere in the direction of its north pole. In the asymptotic regime $\tau \gg \gamma^{-1}$, $\rho_\tau$ converges to the incoherent pure state $|0\rangle\langle 0|$, characterized by a vanishing SME and $\lambda_{\text{min}}(\rho_\tau)\to 0$. In such a regime, the smallest eigenvalue asymptotically behaves as ${\lambda_{\text{min}}}({\rho_{\tau}}) \approx {\lambda_{\text{min}}}(|0\rangle\langle 0|) \approx 0$, leading the weight function ${g_{q}}[{\lambda_{\text{min}}}({\rho_{\tau}})]$ to diverge for any $q \in (0,1)$. Therefore, in the long-time limit, the QSL bound becomes trivial.
The fact that $\tau^{\rm QSL}_{q,z}\to 0$ indicates that, as the system approaches the steady state $|0\rangle\langle 0|$, the rate of information--theoretic change becomes negligible and entropy variations effectively saturate. Hence the speed--limit bound becomes loose (trivial). Although the dynamics do not stop, the further distinguishability gained per unit time vanishes near the fixed point. Consistently, no finite--time orthogonalization occurs: for a qubit under amplitude damping with Bloch vectors $\mathbf r_0$ and $\mathbf r_t$, $F(\rho_0,\rho_t)>0$ for all finite $t$, so the Bures angle never reaches $\pi/2$.

In addition, as visible in Fig.~\ref{figchannel1}, the QSL time $\tau^{\rm QSL}_{q,z}$ exhibits a genuine $q$--dependence: the parameter $q$ tunes the SME sensitivity to the spectrum, so the tightest bound typically arises at an ``optimal'' $q$ set by the dynamics and the initial state. This is complementary to Bures or fidelity-based QSLs \cite{PhysRevX6021031} and is well suited for nonunitary dissipative evolutions.

\textbf{Remark.} Our derivation does not assume CP--divisibility. In non--Markovian dynamics the two--parameter family $\mathcal{E}_{t,s}$ may fail to be CPTP, while the single--parameter map $\mathcal{E}_{t,0}$ remains CPTP and differentiable at each $t$. Hence Eqs.~\eqref{channel1}--\eqref{channel4} still apply with the
time--dependent Kraus operators $V_\ell(t)$. If one prefers a master--equation description, the same bound follows by replacing the Kraus speed with the time--local generator,
$$\left\|\dot\rho_t\right\|_1=\left\|\mathcal{L}_t(\rho_t)\right\|_1,$$
$$\tau \ge
\frac{\big|S_{q,z}(\rho_\tau)-S_{q,z}(\rho_0)\big|}
{2\big\langle g_q[\lambda_{\min}(\rho_t)]\,\|\mathcal{L}_t(\rho_t)\|_1\big\rangle_\tau}.
$$
Memory backflow typically increases either the instantaneous distinguishability or the speed $\|\dot\rho_t\|_1$, which makes the QSL tighter and may induce revivals in the bound.

\subsection{QSL of Dissipative Dynamics Induced by Non-Hermitian Generators}
\label{subsectionNon-Hermitian}
Dissipative quantum systems governed by effective non-Hermitian Hamiltonians have attracted substantial attention in both theoretical studies and experimental researches
~\cite{breuer2002theory, PhysRevA.55.2290, PhysRevLett.111.010402}, together with the QSL in the context of non-Hermitian dynamics \cite{PhysRevLett.123.180403,PhysRevLett.127.100404}.



Building upon the Theorem~\ref{theorem1}, Definition~\ref{definitionerror} and Theorem~\ref{theoremQSL}, we investigate the interplay between the SME and the QSL in the context of quantum systems undergoing nonunitary evolution,
${\mathcal{E}}_t(\cdot) = e^{-i t \text{H}_{\text{sys}}} \, \cdot \, e^{i t  {\text{H}^{\dagger}_{\text{sys}}}}$, where the general effective non-Hermitian Hamiltonian $\text{H}_{\text{sys}}$ can be expressed as
$\text{H}_{\text{sys}} = \text{H}_{\text{Re}} + i\Upsilon_{\text{Im}}$ with
$\text{H}_{\text{Re}}= \frac{1}{2}(\text{H}_{\text{sys}} + {\text{H}^{\dagger}_{\text{sys}}})$  and
$\Upsilon_{\text{Im}}= \frac{1}{2i}(\text{H}_{\text{sys}} - {\text{H}^{\dagger}_{\text{sys}}})$.
Following Refs.~\cite{PhysRevA.106.012403,Int. J. Mod. Phys. B2013,Zlos2015}, we adopt the standard normalized--density approach to effective non-Hermitian generators and work with the physical state,
\[
\rho_t=\frac{\mathcal{E}_t(\rho_0)}{{\rm tr}\!\left[\mathcal{E}_t(\rho_0)\right]},
\]
which guarantees ${\rm tr}(\rho_t)=1$ at all times. Differentiating this definition yields the nonlinear master equation,
\begin{equation}
\frac{d\rho_t}{dt}
= -i[H_{\rm Re},\rho_t]+\{\Upsilon_{\rm Im},\rho_t\}
  -2\,\langle\!\langle \Upsilon_{\rm Im}\rangle\!\rangle_t\,\rho_t,
\label{NonHermitian1}
\end{equation}
with $\langle\langle \cdot\rangle\rangle_t={\rm tr}(\cdot\,\rho_t)$ the instantaneous expectation value.
For an $\eta$--pseudo--Hermitian or $\mathcal{PT}$--symmetric formulation \cite{PRAOhlsson2021,CJPScolarici2006},  one evolves the $\eta$--normalized state $\tilde\rho_t=\eta^{1/2}\rho_t\,\eta^{1/2}/\mathrm{tr}(\eta\rho_t)$
under the $\eta$--inner product. This procedure yields a master equation with the same structure as ours, expressed through $\eta$-expectation values.

Within this framework, we have
\begin{align}
\label{NonHermitian2}
&{\left\| \frac{d{\rho_t}}{dt} \right\|_1}= \|-i [\text{H}_{\text{Re}},{\rho_t}] +\{\Upsilon_{\text{Im}},{\rho_t} \}+ 2{\langle\langle{\Upsilon_{\text{Im}}}\rangle\rangle_t}~{\rho_t}\|_1 \nonumber\\
&\leq \|-i [\text{H}_{\text{Re}},{\rho_t}]\|_1 + \|\{{\Upsilon_{\text{Im}}},{\rho_t} \}\|_1+ \| 2{\langle\langle{\Upsilon_{\text{Im}}}\rangle\rangle_t}\,{\rho_t}\|_1 \nonumber\\
& \leq  2{\| \text{H}_{\text{Re}} \|_{\infty}}\|{\rho_t} \|_1 + 2{\| \Upsilon_{\text{Im}} \|_1}\|{\rho_t} \|_1
+ 2{|\text{tr}(\Upsilon_{\text{Im}}\rho_t)|} ~\|{\rho_t} \|_1\nonumber\\
& \leq  2{\|\text{H}_{\text{Re}} \|_{\infty}}\|{\rho_t} \|_1 + 2{\| \Upsilon_{\text{Im}} \|_1}\|{\rho_t} \|_1
+ 2{\| \Upsilon_{\text{Im}} \|_{\infty}} \|{\rho_t} \|_1~\|{\rho_t} \|_1\nonumber\\
& \leq  2({\| \text{H}_{\text{Re}} \|_{\infty}} + {\| \Upsilon_{\text{Im}} \|_1} + {\| \Upsilon_{\text{Im}} \|_{\infty}}),
\end{align}
where the first inequality follows from the triangle inequality of the trace norm, ${\|A + B\|_1} \leq {\|A\|_1} + {\|B\|_1}$. The second inequality is obtained by invoking the operator norm bounds ${\| [A, B] \|_1} \leq 2 {\|A\|_{\infty}} {\|B\|_1}$ and ${\| AB \|_1} \leq {\|A\|_1} {\|B\|_1}$. The third inequality relies on the trace inequality $|\text{tr}(AB)| \leq {\|A\|_{\infty}} {\|B\|_1}$~\cite{R.Bathia1997,j.laa.2015.07.009.}.
Additionally, we have used the fact that ${\| \rho_t \|_1} = 1$.

Combining Theorem~\ref{theoremQSL} and \eqref{NonHermitian2}, we have the following theorem about the QSL associated with non-Hermitian dynamics.

\begin{theorem}\label{NonHermitianQSL}
The QSL time $\tau_{q,z}^{\mathrm{QSL}}$ satisfies the following bound under the nonunitary dynamics $\mathcal{E}_t(\cdot) = e^{-it \text{H}_{\text{sys}}}\, \cdot \, e^{+it \text{H}^\dagger_{\text{sys}}}$,
\begin{align}
\label{NonHermitian3}
\tau_{q,z}^{\mathrm{QSL}} = \frac{ \left| \text{S}_{q,z}(\rho_\tau) - \text{S}_{q,z}(\rho_0) \right| }{2 \left\langle g_q(\rho_t) \left( \|\text{H}_{\text{Re}} \|_{\infty} + \|\Upsilon_{\text{Im}}\|_1 + \|\Upsilon_{\text{Im}}\|_{\infty} \right) \right\rangle_\tau },
\end{align}
where $\langle \cdot \rangle_\tau$ denotes the time average over the time interval $[0,\tau]$.
\end{theorem}

Analogous to Definition~\ref{definitionerror}, the normalized relative error associated with Theorem~\ref{NonHermitianQSL} is given by
\begin{equation}
\label{NonHermitianerror}
\varsigma_{q,z}(\tau) = 1 - \frac{\left| \text{S}_{q,z}(\rho_{\tau}) - \text{S}_{q,z}(\rho_0)
\right|}{2\left( \| \text{H}_{\text{Re}} \|_{\infty} + \| \Upsilon_{\text{Im}} \|_1 + \| \Upsilon_{\text{Im}} \|_{\infty} \right) \int_0^{\tau} dt \, g_q(\rho_t)}.
\end{equation}
It is evident that, beyond the minimal eigenvalue of  $\rho_t$, the lower bound obtained in Theorem~\ref{NonHermitianQSL} is influenced by the Hermitian components $\text{H}_{\text{Re}}$ and $\Upsilon_{\text{Im}}$, along with the SMEs of both the initial state $\rho_0$ and the final state $\rho_{\tau}$. Notably, at the Hermitian limit that the effective non-Hermitian Hamiltonian reduces to a purely Hermitian operator $\text{H}_{\text{sys}} = \text{H}_{\text{Re}}$ and $\Upsilon_{\text{Im}}$ vanishes, the QSL time in \eqref{NonHermitian3} becomes trivial. This holds for all values of $q$ and $z$ such that $0 < q < 1$ and $0 < z < 1$. This behavior can be attributed to the invariance of the SME under unitary evolution. Specifically, when $\Upsilon_{\text{Im}} = 0$, the entropy remains constant during the evolution, that is,
\[
[{{\text{S}}_{q,z}}({\rho_t})]_{\Upsilon_{\text{Im}} = 0} = {{\text{S}}_{q,z}}(e^{- i t \text{H}_{\text{Re}}} \rho_0 e^{i t \text{H}_{\text{Re}}}) = {{\text{S}}_{q,z}}(\rho_0)
\]
for all $t \in [0, \tau]$. Consequently, this leads to the trivial lower bound $\tau \geq 0$.


It is worth noting that the QSL time formulated in \eqref{NonHermitian3} differs fundamentally from the expression reported in \cite{PhysRevLett.110.050403}. In contrast to our approach, the bound presented in \cite{PhysRevLett.110.050403} is based on the relative purity between $\rho_0$ and $\rho_\tau$ and is inversely related to the variances of both the real $\text{H}_{\text{Re}}$ and imaginary $\Upsilon_{\text{Im}}$.

Furthermore, it should be highlighted that the bound developed in \cite{PhysRevLett.123.180403} provides a broader framework, endowed with a clear geometric perspective grounded in the system's geometric phase. Notably, this generalized formulation has been demonstrated to surpass the performance of both the MT and ML bounds in specific dynamical regimes.


Next we tailor the general results of Sec.~\ref{subsectionNon-Hermitian} to a specific class of non-Hermitian systems governed by the \textit{parity-time ($\mathcal{P}\mathcal{T}$) symmetric Hamiltonians}. Consider the effective Hamiltonian $\text{H}_{\text{sys}}  = \omega{\sigma_x} + i\eta{\sigma_z},$ where $\omega,\eta \in \mathbb{R}$ represent the coupling and dissipation strengths~\cite{10.1093ptepptaa181}. This model features three regimes: (i) unbroken symmetry for $\eta/\omega < 1$ with real spectrum; (ii) criticality at $\eta = \omega$ (exceptional point); and (iii) broken symmetry when $\eta/\omega > 1$, yielding purely imaginary eigenvalues.


Let the system be initially in the state \eqref{ini_state}. The evolved state is given by
$$
{\rho_t} = \frac{e^{-it\text{H}_{\text{sys}}}\,{\rho_0} \, e^{+it \text{H}^\dagger_{\text{sys}}}}{\text{tr}({e^{-it\text{H}_{\text{sys}}}\,{\rho_0} \, e^{+it \text{H}^\dagger_{\text{sys}}}})},
$$
see Appendix~\textbf{A} for the detailed derivations.

Fig.~\ref{figchannel2} illustrates the behavior of the QSL time $\tau^{\,\text{QSL}}_{q,z}$ \eqref{NonHermitian3} versus the dimensionless parameter $\omega\tau$ for the cases $z = 3/4$, $1/2$ and $1/4$. Similar qualitative features are observed for the limits corresponding to the R\'{e}nyi ($z\to 1$) and Tsallis ($z\to q$) entropies.

\begin{figure*}[t]
\centering
\newcommand{\colw}{.33\textwidth}
\newcommand{\gap}{.03\textwidth}
\newcommand{\panel}[2]{%
  \begin{minipage}[t]{\colw}\centering
    \includegraphics[width=\linewidth,trim=6 2 9 1,clip]{#1}\\[-0.2ex]{\small (#2)}
  \end{minipage}%
}
\makebox[\textwidth][c]{%
  \panel{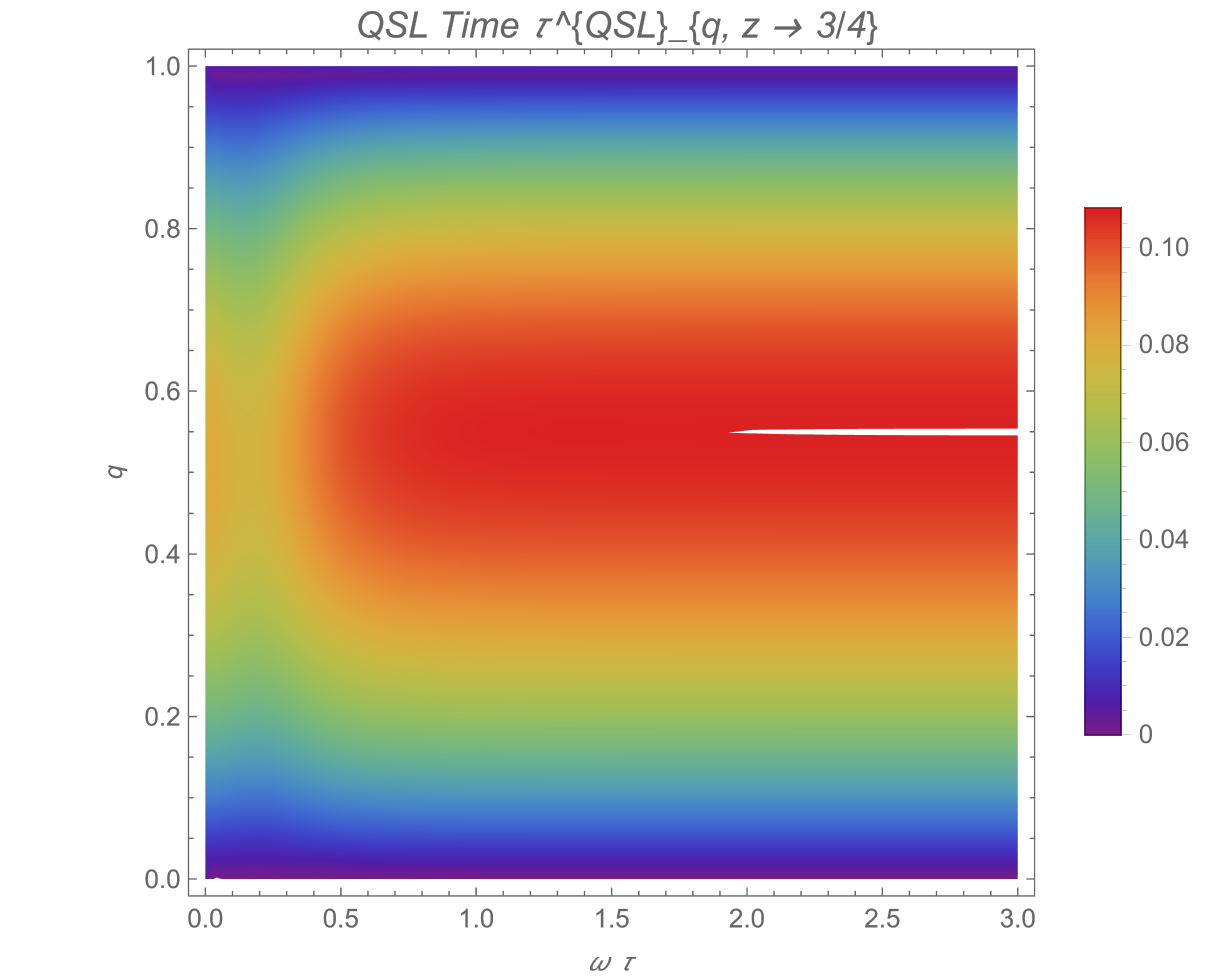}{a}\hspace{\gap}%
  \panel{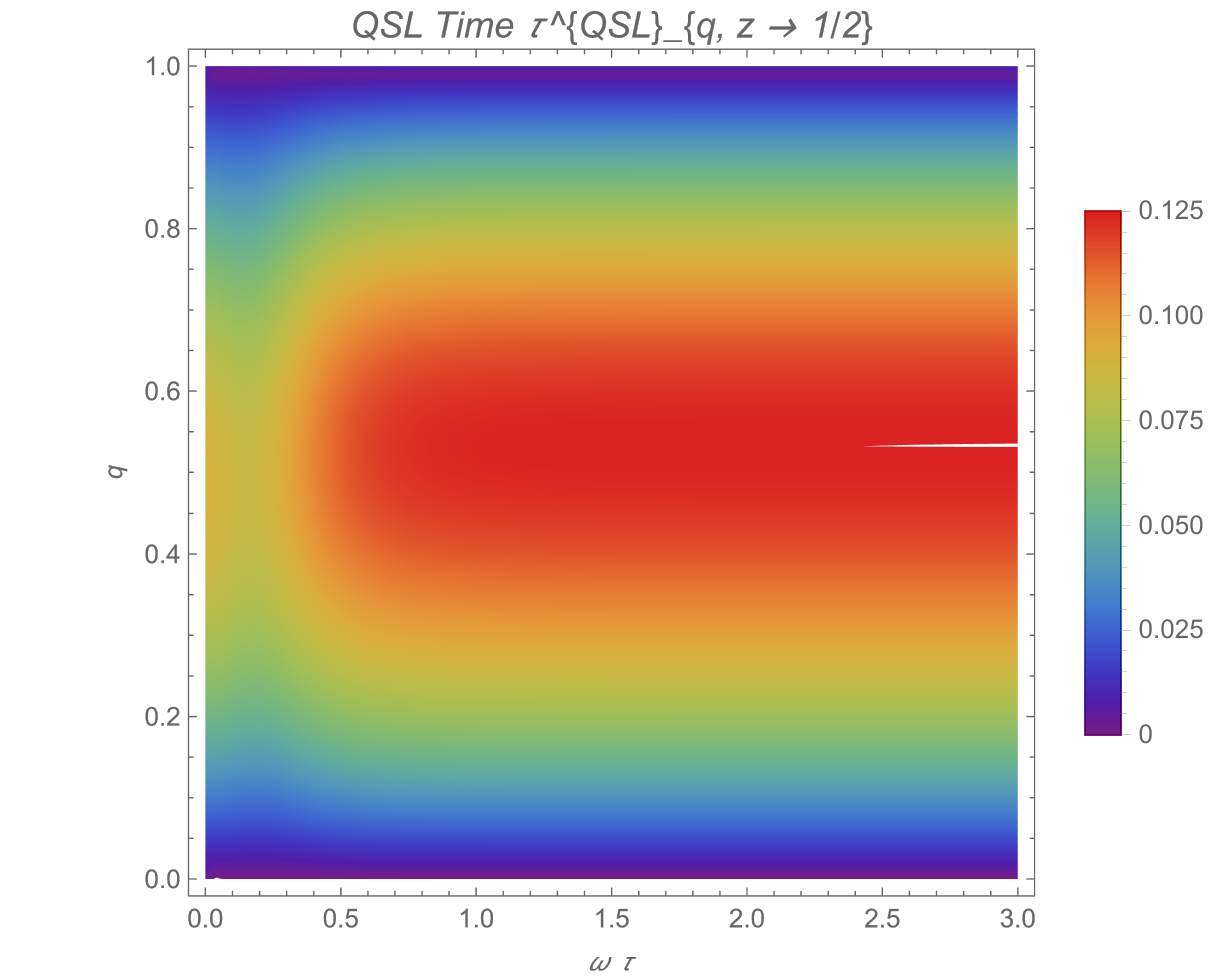}{b}\hspace{\gap}%
  \panel{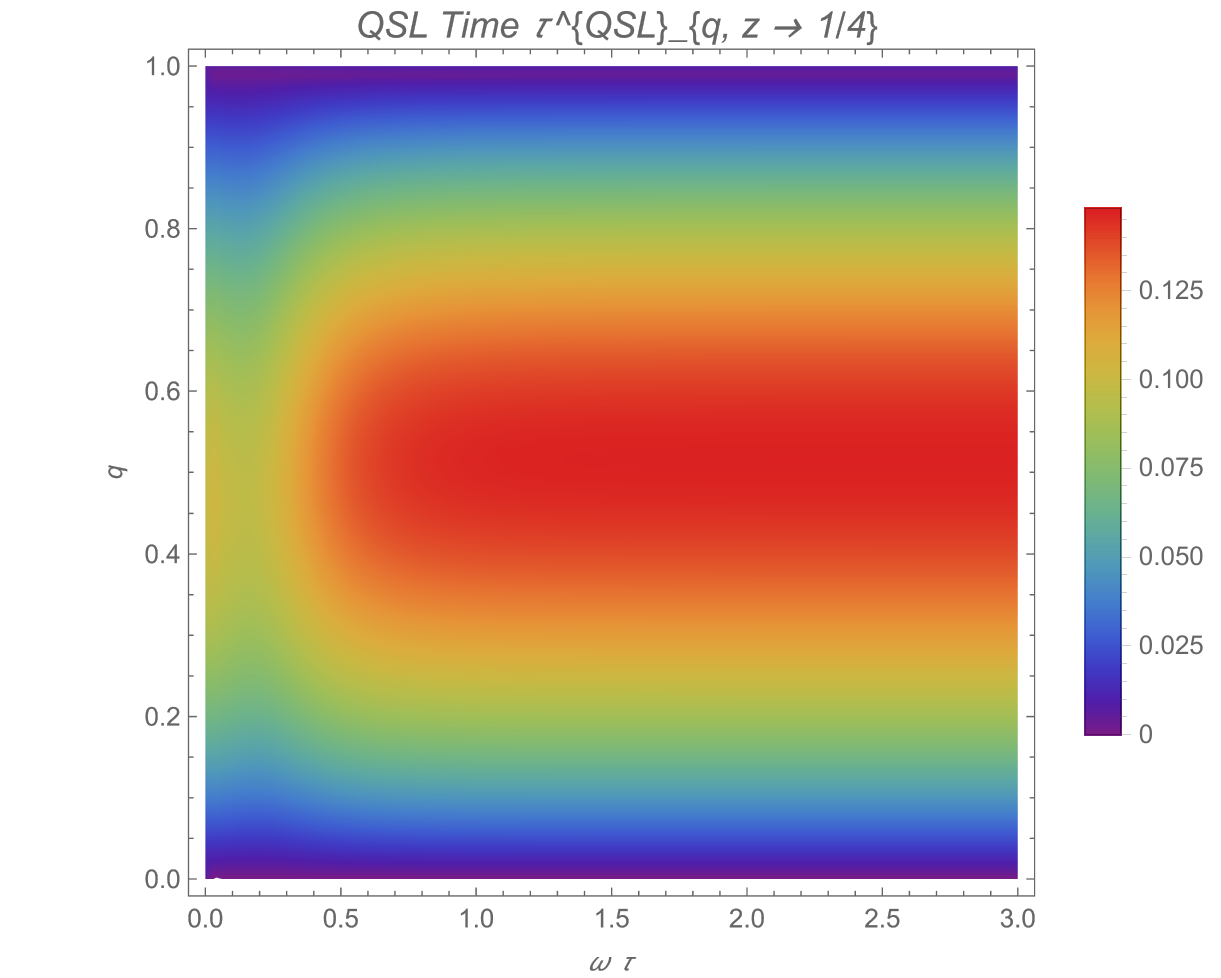}{c}%
}
\vspace{0.9ex}
\makebox[\textwidth][c]{%
  \panel{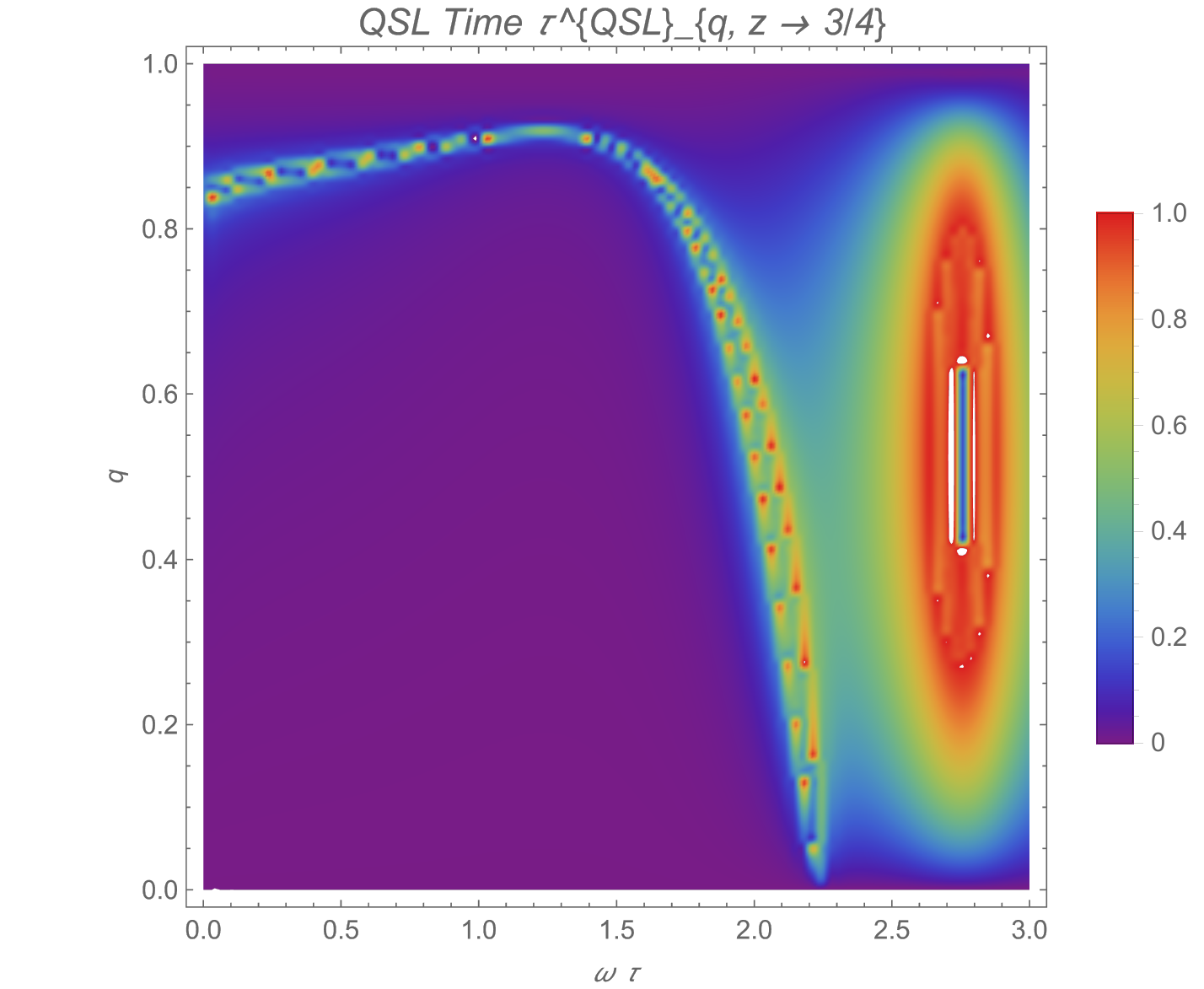}{d}\hspace{\gap}%
  \panel{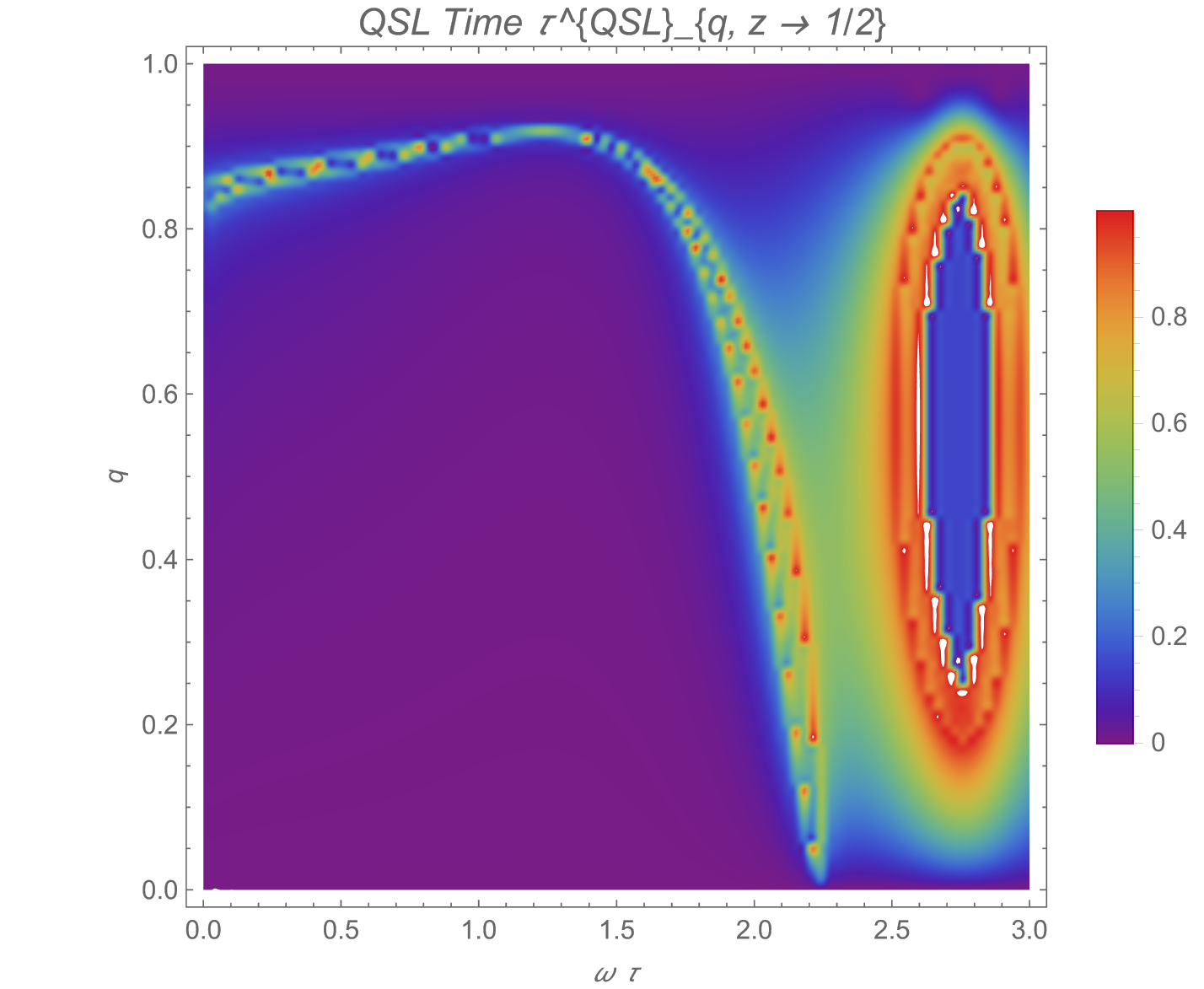}{e}\hspace{\gap}%
  \panel{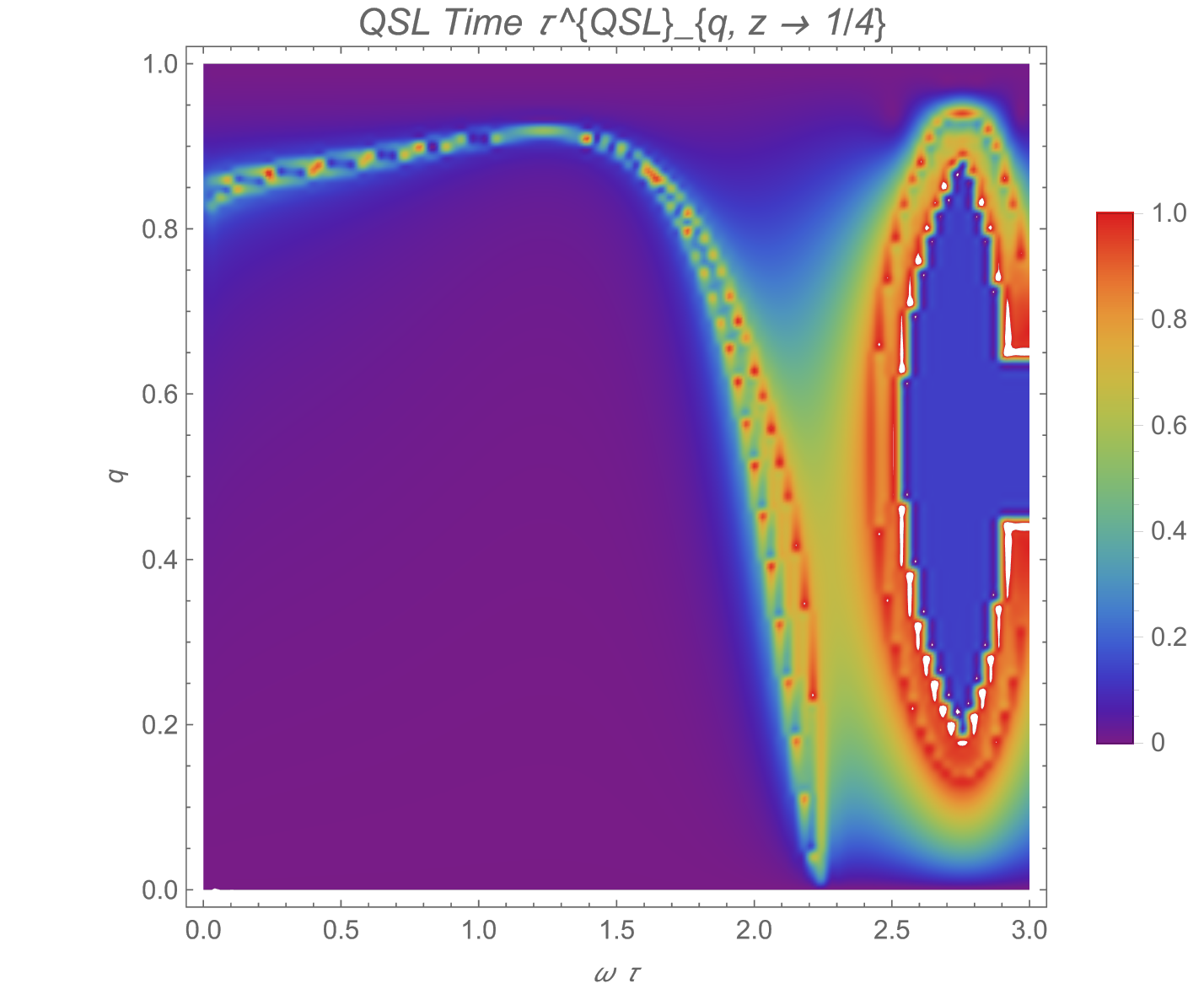}{f}%
  }
\caption{Density plots of the QSL time $\tau^{\text{QSL}}_{q,z}$ \eqref{NonHermitian3} for $z = 3/4$, $1/2$ and $1/4$, for a two-level quantum system evolving under the non-Hermitian Hamiltonian $H{\text{sys}} = \omega \sigma_x + i \eta \sigma_z$. The initial state is given by \eqref{ini_state} with the Bloch vector $\{r,\theta,\phi\}=\{\frac{1}{2},\frac{\pi}{4},\frac{\pi}{4}\}$. $\eta/\omega = 2$ for panels (a)-(c) and $\eta/\omega = 1/2$ for panels (d)-(f).}
\label{figchannel2}
\end{figure*}


Figs.~\ref{figchannel2}(a)-\ref{figchannel2}(c) depict the QSL time in the $\mathcal{P}\mathcal{T}$ broken symmetry phase with $\eta = 2\omega$. In this specific regime, the QSL time mainly displays non-zero values in the range $0.2 \lesssim q \lesssim 0.8$. It reaches its peak in the intervals $0.5 \lesssim \omega \tau \lesssim 3$ and $0.3 \lesssim q \lesssim 0.7$. Moreover, the plots are symmetric about $q = 0.5$. Outside these regions, the QSL time tends to vanish, indicating that the corresponding QSL bounds become loose. Consistently, the relative error shown in  Eq.~\eqref{NonHermitianerror} approaches unity in these regions.

In Figs.~\ref{figchannel2}(d)-\ref{figchannel2}(f), we present the QSL time for the $\mathcal{P}\mathcal{T}$ unbroken symmetry phase with $\eta = \frac{1}{2}\omega$. In this case, the QSL time initially increases with $\omega\tau$, showing nonzero values across $q \in (0,1)$. At early times, apart from a small region where the QSL time remains finite, it rapidly decays and stabilizes near zero, indicating that the QSL bound is loose during this stage. However, at later time, specifically within $2.3 \lesssim \omega\tau \lesssim 3$, the QSL time becomes nonzero again, suggesting that the bound regains significance in this interval.

By comparing Figs.~\ref{figchannel2}(d)-\ref{figchannel2}(f) with Figs.~\ref{figchannel2}(a)-\ref{figchannel2}(c), it becomes evident that overall the QSL bound derived from \eqref{NonHermitian3} exhibits limited tightness across the dynamics. Thus, while the QSL bound constructed from SMEs provides meaningful insights into certain dynamical regimes, its overall looseness highlights the need for further refinement.
In the unbroken $\mathcal{PT}$ phase, the trace--renormalized evolution generated by $H_{\rm sys}$ can produce quasi--returns of the normalized state. Consequently, for a finite patch of $(q,\omega\tau)$ around $\omega\tau\approx 3$ one has $S_{q,z}(\rho_t)\approx S_{q,z}(\rho_0)$. Therefore, the numerator $\big|S_{q,z}(\rho_t)-S_{q,z}(\rho_0)\big|$ in Eq.~\eqref{NonHermitian3} becomes small and the QSL bound turns loose (blue region). Away from the region $\omega\tau\approx 3$ the entropy increases smoothly, yielding the surrounding annular ridge (red ring). This feature does not signal sudden loss or recovery of orthogonality: using the analytical $\rho_t$ from Appendix~A and the qubit identity $F(\rho_0,\rho_t)=\mathrm{Tr}(\rho_0\rho_t)+2\sqrt{\det\rho_0\,\det\rho_t}$ (with $\{r,\theta,\phi\}=\{\tfrac12,\tfrac{\pi}{4},\tfrac{\pi}{4}\}$ and $\eta/\omega\in\{2,\,1/2\}$), one sees that the normalized state stays full rank ($0<\lambda_{\min}(\rho_t)<1$). Hence, $F(\rho_0,\rho_t)>0$ and the Bures angle never reaches $\pi/2$ across the panels. The ring--like patterns therefore stem from entropy variation and the speed term in Eq.~\eqref{NonHermitian3}, not from the changes of orthogonality.

We explore next the potential improvements and discuss the broader implications of our results for nonunitary quantum dynamics. We turn to refine the Schatten speed bound established in \eqref{NonHermitian2}. It is important to highlight that this bound is independent of the evolution time $t$, and that all three terms on its right-hand side rely solely on the Hermitian components of the system's non-Hermitian Hamiltonian, without explicitly depending on the instantaneous quantum state $\rho_t$. To achieve a tighter and more state-sensitive estimation, we modify the first term to incorporate dependence on the quantum state. Specifically, we replace the term $\|-i[H, \rho_t]\|_1$ in \eqref{NonHermitian2} with the variance of the Hermitian operator $H$ evaluated at the initial state $\rho_0$. This replacement is backed up by the subsequent lemma.
\begin{lemma}
\label{Lemma1}
For evolution of the initial state $\rho_0$ to final state $\rho_t$ under the nonunitary dynamics $\mathcal{E}_t(\cdot) = e^{-it \text{H}_{\text{sys}}}\, \cdot \, e^{+it \text{H}^\dagger_{\text{sys}}}$ with $\text{H}_{\text{sys}} = \text{H}_{\text{Re}} + i\Upsilon_{\text{Im}}$, we have
\begin{align}
\label{LemmaA}
\|-i[\text{H}_{\text{Re}} , \rho_t]\|_1  \leq 2 \sqrt{F(\rho_0)} \leq 2 \sqrt{\Delta \text{H}_{\text{Re}}^2 },
\end{align}
where $F(\rho_0)$ is the quantum Fisher information (QFI) of $\rho_0$,
the variance of the operator $\text{H}_{\text{Re}} $  corresponding to  $\rho_0$
is given by $\Delta \text{H}_{\text{Re}}^2 = \text{tr}(\rho_0 \text{H}_{\text{Re}}^2) - \left[\text{tr}(\rho_0 \text{H}_{\text{Re}} )\right]^2$.
\end{lemma}

$\mathit{Proof.}$
Since the Schatten $1$-norm is invariant under unitary transformations, it follows that
\[
\|-i[\text{H}_{\text{Re}} , \rho_t]\|_1 = \|[\text{H}_{\text{Re}} , \mathcal{E}_t(\rho_0)]\|_1 = \|[\text{H}_{\text{Re}}, \rho_0]\|_1.
\]
It has been shown that $\|[\text{H}_{\text{Re}}, \rho_0]\|_1^2 \leq 4 F(\rho_0)$~\cite{PhysRevA.97.022109}. Moreover, the QFI admits an upper bound $F(\rho_0) \leq \Delta \text{H}_{\text{Re}}^2$~\cite{PhysRevA.87.032324}, where equality is attained if the initial state $\rho_0$ is pure. Combining these results, we obtain \eqref{LemmaA}. $\Box$

By using Lemma~\ref{Lemma1}, the Schatten speed bound given in \eqref{NonHermitian2} can be tightened,
\begin{align}
\label{Schattenspeed2}
\left\| \frac{d \rho_t}{dt} \right\|_1 \leq 2 \left( \sqrt{\Delta \text{H}_{\text{Re}}^2} + \| \Upsilon_{\text{Im}} \|_1 + \| \Upsilon_{\text{Im}} \|_{\infty} \right).
\end{align}
Thus, under this refined estimation of the Schatten speed $\left\| \frac{d \rho_t}{dt} \right\|_1 $, the QSL time $\tau_{q,z}^{\mathrm{QSL}}$ satisfies the following lower bound,
\begin{align}
\label{NonHermitian4}
\tau_{q,z}^{\mathrm{QSL}} = \frac{\left| \text{S}_{q,z}(\rho_\tau) - \text{S}_{q,z}(\rho_0) \right| }{2 \left( \sqrt{\Delta \text{H}_{\text{Re}}^{2}} + \|\Upsilon_{\text{Im}}\|_1 + \|\Upsilon_{\text{Im}}\|_{\infty} \right)\left\langle g_q(\rho_t)  \right\rangle_\tau},
\end{align}
while the relative error is given by
\begin{equation}
\label{NonHermitianerror2}
{\varsigma_{q,z}}(\tau) = 1 - \frac{\left| {\text{S}_{q,z}}({\rho_{\tau}}) - {\text{S}_{q,z}}({\rho_0}) \right|}{ 2 (\sqrt{\Delta \text{H}_{\text{Re}}^{2}}+ {\| \Upsilon_{\text{Im}} \|_1} + {\| \Upsilon_{\text{Im}} \|_{\infty}})  {\int_0^{\tau}} dt ~{g_q}(\rho_t)}.
\end{equation}


It is evident that the Schatten speed bound \eqref{NonHermitian4} is tighter than \eqref{NonHermitian3}. This enhancement stems from the fact that the upper bound in \eqref{NonHermitian4} explicitly incorporates the initial state \( \rho_0 \). Specifically, the variance term reflects the degree of uncertainty associated with the observable \( H \) in the state \( \rho_0 \), thereby linking the bound to the system's preparation. In contrast, the bound given by \eqref{NonHermitian3} relies solely on the system's Hamiltonian.

\section{Quantum speed limit for Sharma-Mittal entropy in many-body systems}\label{sect4}
The SME has emerged as an invaluable means of quantifying quantum correlations in multipartite systems, as evidenced in \cite{SMRelativeDivergence2015}. Spurred by the discussions detailed in Sec.~\ref{sect3}, we herein embark on an exploration of the link between quantum correlations and QSLs within the framework of nonunitary dynamics which is in charge of governing the subsystems present in expansive many-body quantum systems.


Let us consider a finite-dimensional closed bipartite quantum system ${\mathcal{H}_A} \otimes {\mathcal{H}_B}$. The dimensions of the two subsystems are $d_A$ and $d_B$, respectively. The overall time-invariant Hamiltonian that controls the composite system $AB$ is expressed as
$$
\text{H}_{\text{sys}} =\text{H}_A \otimes \mathbb{I}_B + \mathbb{I}_A \otimes \text{H}_B + \text{H}_{AB},
$$
where $\text{H}_A$ ($\text{H}_B$) acts solely on $\mathcal{H}_A$ ($\mathcal{H}_B$), and $\text{H}_{AB}$ accounts for the interaction between the two subsystems.

Initially, the system is set up in a state $\rho_0$, which may be either a pure quantum state or a mixed one. Subsequently, it evolves according to the global unitary dynamics characterized by
$$
{\mathcal{E}}_t(\rho_0) = e^{-i t \text{H}_{\text{sys}}} \rho_0 e^{i t \text{H}_{\text{sys}}}.
$$
Therefore, the corresponding marginal states are given by
$$
\rho^{A,B}_t = \mathrm{tr}_{B,A}\left({\mathcal{E}}_t(\rho_0)\right),
$$
and the nonunitary dynamics of the reduced states is governed by
\begin{equation}
\label{many-body1}
\frac{d \rho^{A,B}_t}{dt} = -i\, \mathrm{tr}_{B,A}\left( [\text{H}_{\text{sys}}, {\mathcal{E}}_t(\rho_0)] \right).
\end{equation}


Without loss of generality, we proceed to examine the dynamic behavior of the SMEs  ${{\text{S}}_{q,z}}(\rho^A_t)$ associated with the subsystem $A$. According to the results derived in Sec.~\ref{sect2}, the $\left| \frac{d}{dt}{{\text{S}}_{q,z}}(\rho^A_t) \right|$ is subject to an upper bound
\begin{equation}
\label{many-body2}
\left| \frac{d}{dt}{{\text{S}}_{q,z}}(\rho^A_t) \right| \leq {g_{q}}[{\lambda_{\text{min}}}(\rho^A_t)] \left\| \frac{d \rho^A_t}{dt} \right\|_1,
\end{equation}
where $g_q$ is given in \eqref{g_q}.

For the Schatten speed associated with the $\rho^A_t$, we have the following lemma.

\begin{lemma}
\label{Lemma2}
The reduced state $\rho^A_t$ obeys the inequality below,
\begin{equation}
\label{many-body4}
\left\| \frac{d \rho^A_t}{dt} \right\|_1 \leq \|[\text{H}_{\text{sys}}, \rho_0]\|_1 \leq 2 \sqrt{F(\rho_0)} \leq 2 \sqrt{\Delta \text{H}_{\text{sys}}^2},
\end{equation}
where the variance of $\text{H}_{\text{sys}}$ with respect to $\rho_0$ is given by
$\Delta \text{H}_{\text{sys}}^2 = \mathrm{tr}(\rho_0  \text{H}_{\text{sys}}^2) - \left( \mathrm{tr}(\rho_0  \text{H}_{\text{sys}}) \right)^2$.
\end{lemma}

$\mathit{Proof.}$
For an operator $\mathcal{X} \subset {\mathcal{H}_A} \otimes {\mathcal{H}_B}$, it has been shown in \cite{PhysRevA.78.012308,Rastegin2012} that
$\| \mathrm{tr}_B(\mathcal{X}) \|_p \leq d_B^{(p-1)/p} \| \mathcal{X} \|_p$.
Setting $p=1$ and choosing $\mathcal{X} = -i [ \text{H}_{\text{sys}}, \mathcal{E}_t(\rho_0)]$, we obtain
\[
\left\| \frac{d \rho^A_t}{dt} \right\|_1 \leq \| [ \text{H}_{\text{sys}}, \rho_0] \|_1.
\]
By using Lemma~\ref{Lemma1}, we complete the proof.
$\Box$

Using Lemma~\ref{Lemma2}, we can refine the bound on $\left| \frac{d}{dt} \text{S}_{q,z}(\rho^A_t) \right|$ in \eqref{many-body2},
\begin{equation}
\label{many-body3}
\left| \frac{d}{dt} \text{S}_{q,z}(\rho^A_t) \right| \leq 2 \, {g_{q}}[{\lambda_{\text{min}}}(\rho^A_t)]  \, \sqrt{\Delta  \text{H}_{\text{sys}}^2}.
\end{equation}
Subsequently, by taking into account the property that $\text{H}_{\text{sys}}$ is time-invariant, the integration of \eqref{many-body3} over $t\in [0,\tau]$ gives rise to
\begin{equation}
\label{many-body5}
\left| {{\text{S}}_{q,z}}({\rho^A_{\tau}}) -  {{\text{S}}_{q,z}}({\rho^A_0}) \right| \leq 2\, \tau \sqrt{\Delta  \text{H}_{\text{sys}}^2}\,{\left\langle {g_{q}}[{\lambda_{\text{min}}}({\rho^A_t})] \right\rangle_{\tau}},
\end{equation}



It is important to highlight that \eqref{many-body5} establishes an upper bound on the variation of the SME between the initial reduced state $\rho^A_{0}$ and final reduced state $\rho^A_{\tau}$, based on the quantum fluctuations of the system's Hamiltonian. Specifically, this relation shows that the correlations captured by the SME are constrained by $\Delta  \text{H}_{\text{sys}}^2$. The bound also involves the time-averaged contribution of ${\lambda_{\text{min}}}({\rho^A_t})$.

\begin{theorem}
\label{QSLreduced}
For a bipartite state $\rho_0$ evolving under global unitary evolution, $\mathcal{E}_t(\rho_0)=e^{-i\text{H}_{\text{sys}}t}\rho_0e^{i\text{H}_{\text{sys}}t}$,
the QSL time corresponding to the dynamics of $\rho_t^A=\mathrm{tr}_B[\mathcal{E}_t(\rho_0)]$ has the following lower bound,
\begin{equation}
\label{many-body6}
\tau_{q,Z}^{\mathrm{QSL}} := \frac{\left| \text{S}_{q,Z}(\rho_\tau^A) - \text{S}_{q,Z}(\rho_0^A) \right|}{2 \sqrt{\Delta  \text{H}_{\text{sys}}^2}\, \left\langle g_q [{\lambda_{\text{min}}}({\rho^A_t})]  \right\rangle_\tau}~.
\end{equation}
\end{theorem}

To assess the tightness of the bounds given in \eqref{many-body5} and~\eqref{many-body6}, we introduce the corresponding relative error,
\begin{equation}
\label{many-body7}
\varsigma_{q,z}(\tau) = 1 - \frac{\left| {{\text{S}}_{q,z}}({\rho^A_{\tau}}) - {{\text{S}}_{q,z}}({\rho^A_0})  \right|}{ 2\sqrt{\Delta{ \text{H}_{\text{sys}}^2}}  {\int_0^{\tau}} dt \, {g_{q}}[{\lambda_{\text{min}}}({\rho^A_t})]   } ~.
\end{equation}

It is noted that the QSL $\tau^{\text{QSL}}_{q,z}$ exhibits an inverse dependence on the Hamiltonian variance $\Delta{ \text{H}_{\text{sys}}^2}$, in consistent with the MT-type QSL bounds. Furthermore, formulated in terms of the SME, \eqref{many-body7} underscores the influence of quantum correlations on the temporal scale required to transform the initial marginal state $\rho_0^A$ into the entangled configuration $\rho_t^A$ over the interval $t \in [0, \tau]$. In particular, a strictly positive value of $\tau^{\text{QSL}}{q,z}$ signifies the presence of nonzero entanglement generated throughout the subsystem evolution. Conversely, when the correlation signature is negligible, namely, when ${{\text{S}}_{q,z}}(\rho^A_{\tau}) \approx {\text{S}}_{q,z}(\rho^A_0)$, the QSL lower bound vanishes, $\tau^{\text{QSL}}_{q,z} \approx 0$.

In particular, at the limit $q \to 1$ and $z \to 1$, \eqref{many-body7} converges to zero. This demonstrates that in this particular scenario, the von Neumann entropy is associated with a non-informative (trivial) QSL. It is also relevant to point out that, by using the concept of quantum fidelity, Ref.~\cite{PhysRevX.9.011034} addresses QSLs in the context of ground-state evolution under unitary dynamics driven by many-body control Hamiltonians. Here, we adopt a different perspective. Specifically, we employ the SME as an informational measure to explore the nonunitary evolution of the reduced (marginal) states in general quantum many-body systems. Our method applies to arbitrary initial states and does not restrict the analysis to ground-state manifolds alone.


To further illustrate our results, we turn our attention to two paradigmatic quantum many-body systems. We begin by considering the $\textit{the spin-$\frac{1}{2}$ XXZ model}$ defined on a chain of $L$ sites with open boundary conditions. The Hamiltonian of the system is given by
\begin{equation}
 \text{H}_{\text{sys}} = J \sum_{j = 1}^{L - 1} \left( \sigma_j^x \sigma_{j+1}^x + \sigma_j^y \sigma_{j+1}^y + \Delta\, \sigma_j^z \sigma_{j+1}^z \right),
\end{equation}
where $J$ stands for the exchange coupling strength and $\Delta$ signifies the anisotropy \cite{takahashi_1999,giamarchi_2003}.

The system commences with an initial mixed state,
\begin{equation}
\label{XXZstate}
\rho_0 = \frac{1 - p}{d} \mathbb{I} + p\, |\Phi\rangle\langle\Phi|,
\end{equation}
where \(|\Phi\rangle = |1,0,1,0,\ldots,1,0\rangle\) is the N\'{e}el state and \(p\in[0,1]\). For a chain of length \(L\), the dimension \(d = 2^L\). In this context, \(|0\rangle\) and \(|1\rangle\) respectively stand for the spin-up and spin-down states.
The variance of the Hamiltonian of the XXZ model is given by
\[
\Delta  \text{H}_{\text{sys}}^2 = J^2 \left[ \frac{5}{2} + \frac{15}{16} \Delta^2 \right].
\]
Under the thermodynamic limit \( L \to \infty \), the variance exhibits a linear scaling with respect to the size of the system. For \( L=2 \) and \( p=1/4 \), the derivations of the QSL time \eqref{many-body6} are provided in \textbf{Appendix B}.
\begin{figure*}[t]
\centering

\newcommand{\colw}{.333\textwidth}
\newcommand{\gap}{.03\textwidth}
\newcommand{\panel}[2]{%
  \begin{minipage}[t]{\colw}\centering
    \includegraphics[width=\linewidth]{#1}\\[-0.2ex]{\small (#2)}
  \end{minipage}%
}
\makebox[.99\textwidth][c]{%
  \panel{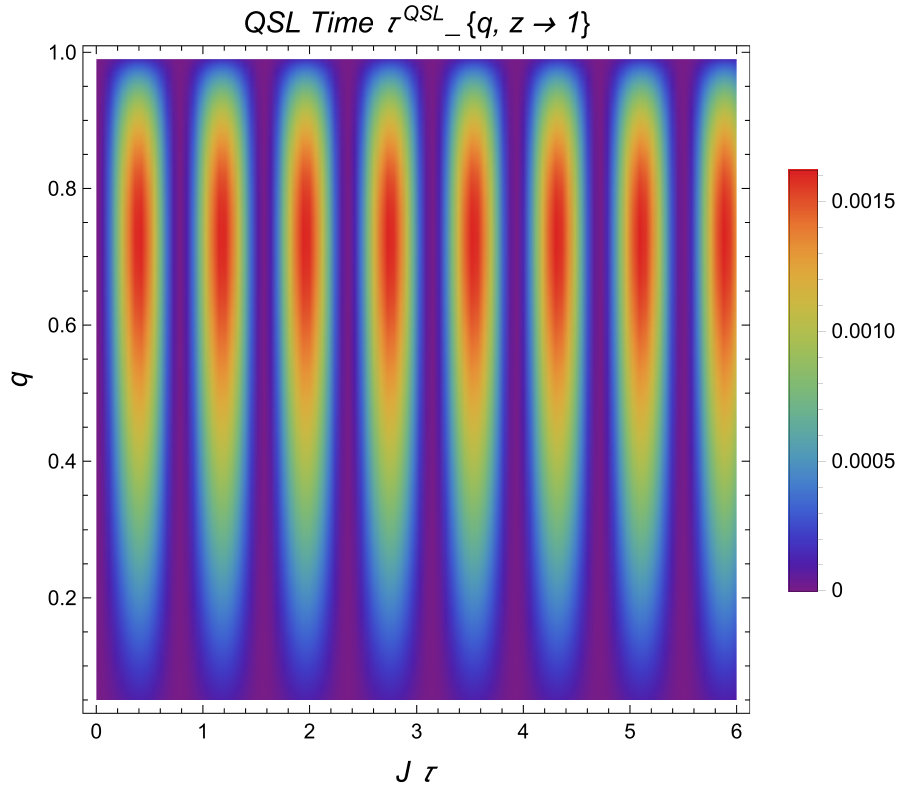}{a}\hspace{\gap}%
  \panel{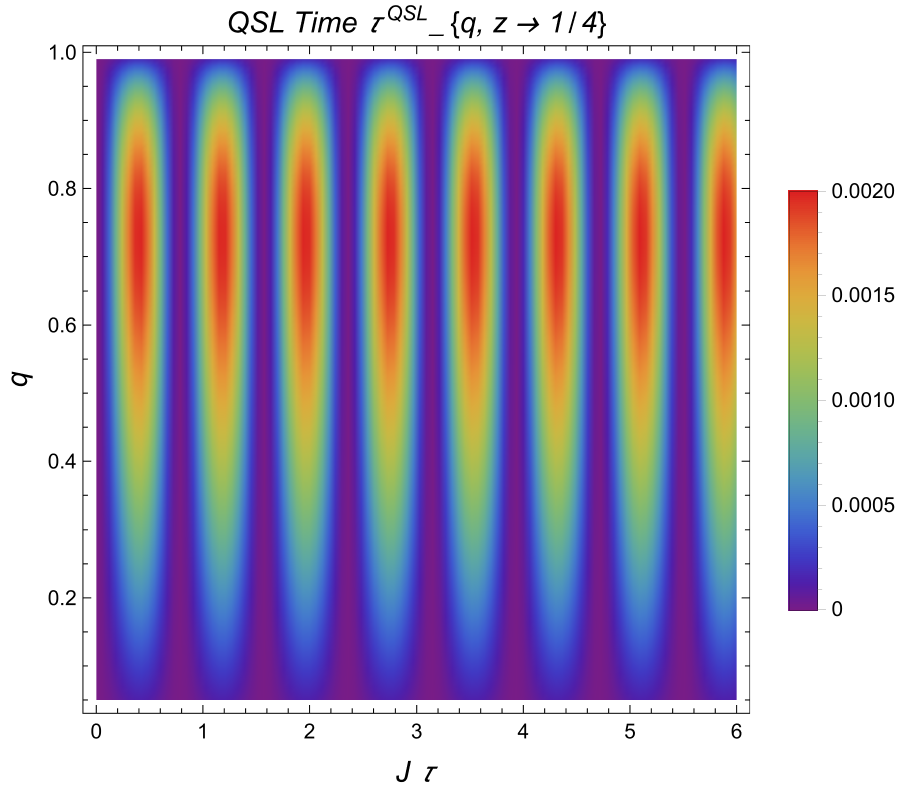}{b}\hspace{\gap}%
  \panel{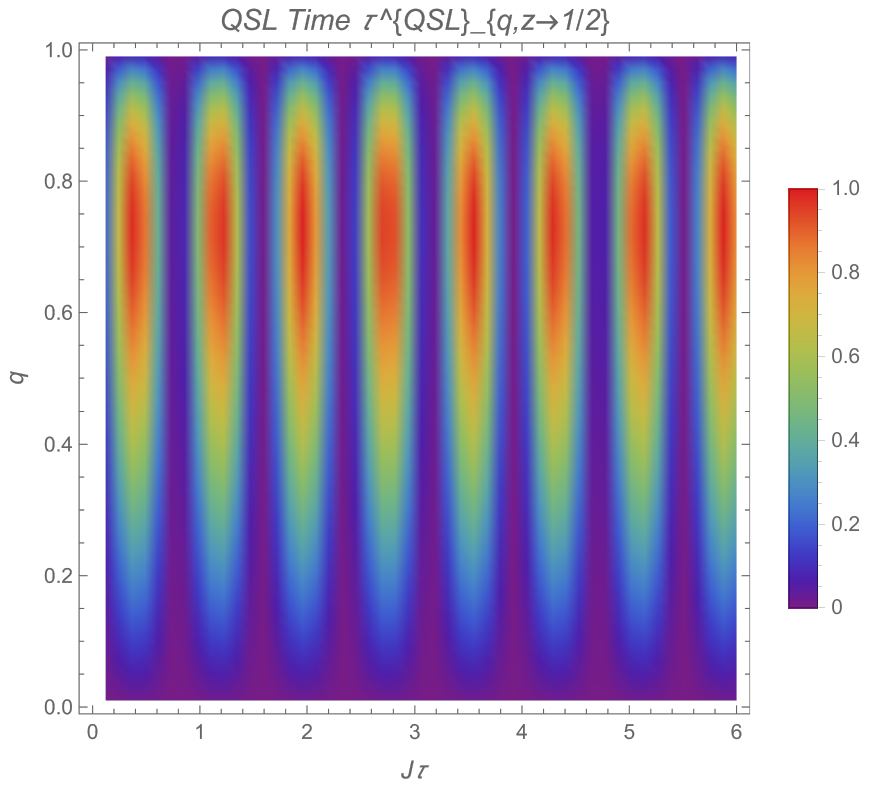}{c}%
}
\vspace{0.9ex}
\makebox[.99\textwidth][c]{%
  \panel{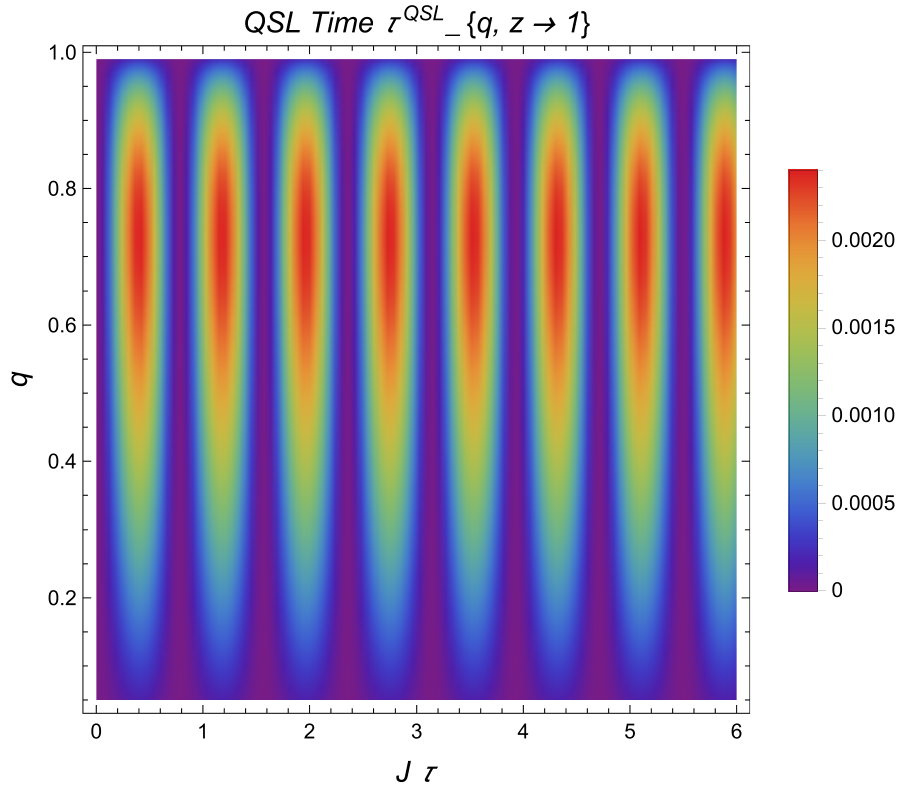}{d}\hspace{\gap}%
  \panel{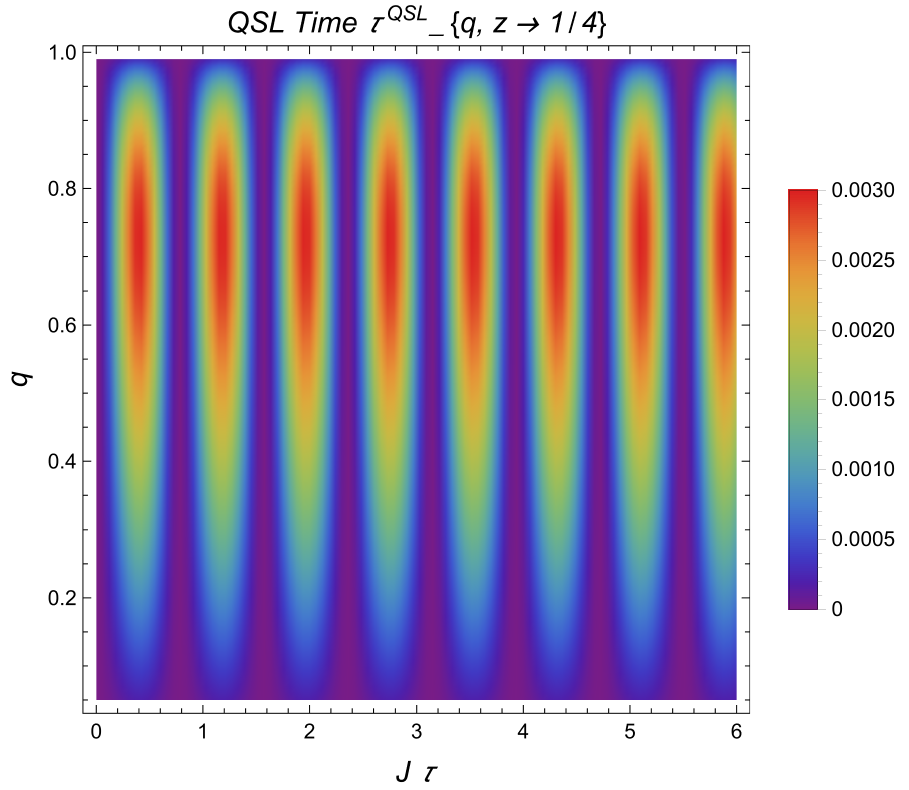}{e}\hspace{\gap}%
  \panel{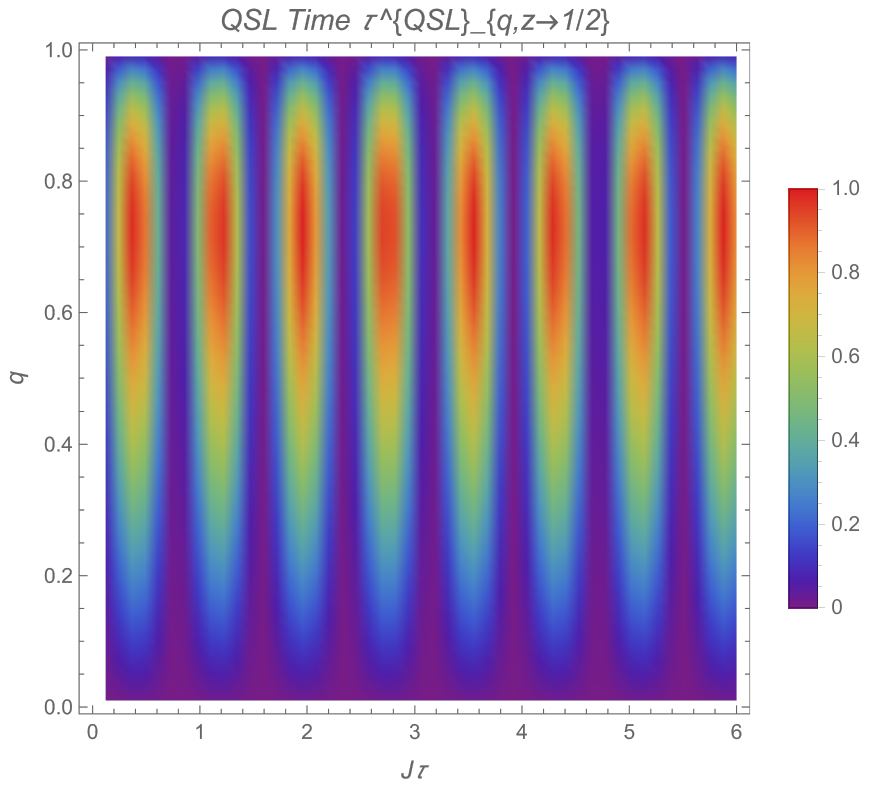}{f}%
}
\caption{%
Density plots of the QSL time $\tau^{\rm QSL}_{q,z}$ [Eq.~(\ref{many-body6})] for the spin-$\tfrac12$ XXZ chain with $J\Delta=0.5$.
The initial state is given by Eq.~(\ref{XXZstate}). The chain length is $L=2$ (open boundary), with a bipartition $L_A=L_B=1$.
The mixing parameter is $p=1/4$.}
\label{figtwoqubit}
\end{figure*}
In Figs.~\ref{figtwoqubit}, we present the QSL time \eqref{many-body6} by considering the R\'{e}nyi entropy $\text{R}_{q} (\rho)= \text{S}_{q,z \to 1}(\rho)$, the Sharma-Mittal entropies $\text{S}_{q,z = 1/4}(\rho)$ $(z=1/4)$ and $\text{S}_{q,z = 1/2}(\rho)$ $(z=1/2)$, pertained to the XXZ model with open boundary conditions. We set the value of the anisotropy parameter such that $J\Delta = 0.5$, and the coupling constant: $J = 1/2$ for Figs.~\ref{figtwoqubit}(a)-\ref{figtwoqubit}(c) and $J = 1/4$ for Figs.~\ref{figtwoqubit}(d)-\ref{figtwoqubit}(f). The system is initialized in a mixed state with $p = 1/4$. The system size is fixed at $L = 2$ with $L_A = 1$ and $L_B = 1$. Here, it is important to note that although the global system $AB$ experiences a unitary evolution governed by the time-independent XXZ Hamiltonian, the resulting reduced dynamics of the subsystem $A$ turns out to be nonunitary.

By examining Figs.~\ref{figtwoqubit}, we note that, across all the cases under consideration, the QSL time displays an oscillatory pattern as it varies with time.
The oscillation features depend strongly on both the coupling strength $J$ and the entropy parameter $z$. Comparing Fig.~\ref{figtwoqubit}(a)--\ref{figtwoqubit}(c) ($J=1/2$) with Fig.~\ref{figtwoqubit}(d)--\ref{figtwoqubit}(f) ($J=1/4$) shows that a larger $J$ leads to faster oscillations and smaller QSL amplitudes, consistent with the analytical period relation $T \propto 1/J$.
For a fixed $J$, the SME further shows that increasing $z$ from $1/4$ to $1/2$ results in smoother QSL evolution with smaller temporal variations.
Among all cases, $z=1/2$ gives the most stable and physically reasonable result, indicating that this choice captures the reduced system's dynamical evolution more accurately.

For $0<q<1$, the QSL shows clear recurrence patterns, punctuated by intervals where the bound approaches (or nearly reaches) zero. This reflects the rich dynamical features of the reduced subsystem evolving under the influence of the global many--body dynamics. Although the global $AB$ dynamics is unitary, the \emph{local} single-site state undergoes nonunitary evolution and remains mixed throughout. For the initial mixture with $p=1/4$ one finds (see \textbf{Appendix B}) that the eigenvalues satisfy $\lambda_{1,2}(t)\in[3/8,\,5/8]$ for all $t$, so the local spectrum never reaches $\{0,1\}$ and the reduced state cannot become orthogonal to its initial reduced state. The observed revivals arise from information backflow between the site and the rest of the chain, which transiently enhances distinguishability or the speed $\|\dot\rho_t\|_1$ and thereby tightens the bound.

In the end, we offer some general observations regarding the characteristics of the QSL time ${\tau^{\text{QSL}}_{q,z}}$ and the relative error $\varsigma({\rho^A_{\tau}})$ within the context of the XXZ model. It is observed that the QSL time is inversely related to the $\Delta  \text{H}_{\text{sys}}^2$, namely, ${\tau^{\text{QSL}}_{q,z}} \sim \delta({\rho^A_{\tau}})/\sqrt{\Delta  \text{H}_{\text{sys}}^2}$, where the function
$\delta({\rho^A_{\tau}}) = |{{\text{S}}_{q,z}}({\rho^A_{\tau}}) - {{\text{S}}_{q,z}}({\rho^A_0})| {\left\langle {g_{q}}[{\lambda_{\text{min}}}({\rho^A_t})]\right\rangle_{\tau}^{-1}}$
depends on the subsystem size $L_A$. It is worthy noting that a smaller relative error corresponds to a tighter QSL bound, as given in \eqref{many-body6}. A similar conclusion has been drawn in previous studies concerning equilibration times in many-body systems, although focusing on relative purity as the distinguishability metric between quantum states~\cite{PhysRevA.104.052223}.

\section{Conclusion}\label{sect5}

We have comprehensively explored the speed limits of finite dimensional quantum systems undergoing arbitrary nonunitary evolutions, based on the the quantum Sharma-Mittal entropy (SME). Our main contribution lies in deriving a family of QSLs associated with the two-parameter SME, applicable to general nonunitary physical processes.

We have analyzed the time derivative of the SME and established an upper bound involving the minimum eigenvalue of the evolved state and the evolved state corresponding to the Schatten speed. In view of this, we have derived a QSL for nonunitary dynamics, showing that the bound explicitly depends on the SME, the minimum eigenvalue and the Schatten speed. The proposed bounds are computationally efficient and encompass several specific cases within the parameter range \( q \in (0,1) \) and \( 1 > z > q > 0 \). We have explicitly characterized the domain of applicability of the QSLs and analyzed various special cases.


These bounds have been then implemented in the context of quantum channels as well as non-Hermitian dynamic processes, illustrated through the amplitude damping channel and $\mathcal{P}\mathcal{T}$-symmetric Hamiltonians. In the context of many-body quantum systems, we have derived an upper constraint for the nonunitary dynamics of marginal states (Sec.~\ref{sect5}), showing that the SME evolution is governed by quantum fluctuations of local multipartite Hamiltonians. Additionally, we have discovered that there exists an inverse relationship between the QSL and the variance of the Hamiltonian. This finding aligns with the characteristics of the MT-type QSLs.


Furthermore, the current framework admits a natural extension to the Sharma-Mittal relative divergence~\cite{SharmaMittalJMathSci1975,SMRelativeDivergence2015}, which may lead to even tighter QSLs. It would be worthwhile to explore possible trade-off relations among SMEs, QSLs and quantum correlations in general physical processes. Consequently, our results may highlight further investigations on quantum dynamics, with potential applications in thermalization in many-body systems, noisy quantum metrology, quantum computing and quantum communication.

\section*{ACKNOWLEDGEMENTS}
This work is supported by the National Natural Science Foundation of China (NSFC) under Grant Nos. 12171044; the specific research fund of the Innovation Platform for Academicians of Hainan Province.

\section*{APPENDIX}

\section*{Appendix A: Exact Solution for the Dynamics of a $\mathcal{PT}$-Symmetric Two-Level System}
We carry out an analysis on the nonunitary evolution regulated by the two-level non-Hermitian Hamiltonian, $H_{\text{sys}} = \mathbf{\vec{u}} \cdot \vec{\sigma}$,
with $\mathbf{\vec{u}} = \{\omega, 0, i\eta\}$, $\omega, \eta \in \mathbb{R}$, and  $\vec{\sigma} = \{\sigma_x, \sigma_y, \sigma_z\}$.
The initial state is a single-qubit density matrix given by \eqref{ini_state}. We fix $\{r, \theta, \phi\} = \left\{ \frac{1}{2}, \frac{\pi}{4}, \frac{\pi}{4} \right\}$.

The time evolution of the state is described by
\begin{equation}
\rho_t = \frac{K_t \rho_0 K_t^\dagger}{\operatorname{tr}(K_t \rho_0 K_t^\dagger)},
\end{equation}
where $K_t = e^{-i t H_{\text{sys}}}$ is the nonunitary evolution operator generated by $H_{\text{sys}}$.
We begin with analyzing the QSL time in the regime where the system exhibits broken $\mathcal{P}\mathcal{T}$ symmetry, characterized by $\eta > \omega$. Specifically, we firstly consider the case $\eta = 2\omega$.
Under these circumstances, the evolution Kraus operator $K_t$ can be expressed as
\begin{equation}
K_t = \cosh(\sqrt{3} \, \omega t) \, \mathbb{I}
+ \sinh(\sqrt{3} \, \omega t) \, (\hat{u} \cdot \vec{\sigma}),
\end{equation}
where the unit vector $\hat{u}$ is defined as
\begin{equation}
\hat{u} = -i \frac{\mathbf{\vec{u}}}{\sqrt{3} \, \omega} = \left\{ \frac{i}{\sqrt{3}}, 0, \frac{2}{\sqrt{3}} \right\},
\end{equation}
by substituting $\mathbf{\vec{u}} = \{\omega, 0, 2i\omega\}$.

As a result, the evolved state takes the form
\begin{align*}
{\rho_t} = \frac{1}{4 + 2{m_t}}\left(4{\rho_0} + {m_t}\mathbb{I} + {\vec{b}_t}\cdot\vec{\sigma}\right),
\end{align*}
where
\begin{align*}
{m_t}= \frac{14}{3}{\sinh^2}( \sqrt{3}\omega t \,)-\sqrt{\frac{2}{3}}{\sinh}(2\sqrt{3}\omega t),
\end{align*}
\begin{widetext}
\begin{align*}
\vec{b}_t = \left\{
\frac{1}{2} - \frac{(1+2\sqrt{2}i)\sinh^2\left( \sqrt{3}\omega t \right)}{3} , \frac{1}{2} - \frac{\sinh\left(2\sqrt{3}\omega t\right)}{\sqrt{6}} + \frac{7\sinh^2\left( \sqrt{3}\omega t \right)}{3},~\frac{\sqrt{2}}{2} - \frac{7\sinh\left(2\sqrt{3}\omega t\right)}{2\sqrt{3}}  + \frac{2(\sqrt{2}i-1)\sinh^2\left( \sqrt{3}\omega t \right)}{3\sqrt{2}}
\right\}.
\end{align*}
\end{widetext}
Hence, the minimal eigenvalue of $\rho_t$  can be expressed explicitly as
\begin{widetext}
\begin{align*}
&\lambda_{\text{min}}(\rho_t)=\\
&-\frac{-4\sqrt{6}A+28 B+\sqrt{2} \sqrt{-4 \sqrt{3} \left(13 \sqrt{2}+3 i\right) C+8 \left(-14 \sqrt{2}+3 i\right) \sqrt{3}A+8 \left(1+16 i \sqrt{2}\right) B+7 \left(55-8 i \sqrt{2}\right) D-72 i \sqrt{2}+111}-4}{8 \left(\sqrt{6}A-7B+1\right)},
\end{align*}
\end{widetext}
where
$A=\sinh \left(2 \sqrt{3} \omega t\right)$, $B=\cosh \left(2 \sqrt{3} \omega t\right)$,
$C=\sinh \left(4 \sqrt{3} \omega t\right)$ and $D=\cosh \left(4 \sqrt{3} \omega t\right)$.

Next, we analyze the QSL time in the $\mathcal{P}\mathcal{T}$ symmetry-preserving regime, corresponding to $\eta < \omega$, by choosing $\eta = \omega/2$.
In this scenario, the evolution operator takes the form
\begin{equation}
K_t = \cos\left( \frac{\sqrt{3} \omega t}{2} \right)\mathbb{I} - i\sin\left( \frac{\sqrt{3} \omega t}{2}\right)(\hat{u}\cdot\vec{\sigma}),
\end{equation}
where $\hat{u} =\left\{ \frac{2}{\sqrt{3}}, 0, \frac{i}{\sqrt{3}} \right\}$ is a unit vector.

Accordingly, the evolved state can be written as
\begin{equation}
\rho_t = \frac{1}{4 + 2m_t} \left(4\rho_0 + m_t \mathbb{I} + \vec{b}_t \cdot \vec{\sigma}\right).
\end{equation}
The corresponding matrix form is given by
\begin{equation}
\rho_t = \frac{1}{2M+4}
\begin{pmatrix}
M+Q+2 & M - i*N + \frac{1}{2} + 2 \\
M + i*N + \frac{1}{2} + 2 & M-Q+2
\end{pmatrix}
\end{equation}
with $a={\sinh}(2\sqrt{3}\omega t)$, $b={\sinh^2}( \sqrt{3}\omega t )$,
$M=\frac{a}{\sqrt{6}}+2 b$,
$N=\frac{4 a^2}{3}-\frac{\sqrt{6} b}{3}+\frac{1}{2}$ and
$Q=\sqrt{2} (-a)+\sqrt{3} b+\frac{\sqrt{2}}{2})$.
Thus, the minimal eigenvalue of the  $\rho_t$ is

\begin{widetext}
\begin{align*}
&\lambda_{\text{min}}(\rho_t)=\\
&\frac{\sqrt{-40 \left(\sqrt{6}-36\right)A-8 \sqrt{6} \sin \left(3 \sqrt{3} \omega t\right)+80 \sqrt{6} B-433 D-20 \left(3 \sqrt{6}+11\right) E-16C+2 \cos \left(4 \sqrt{3} \omega t\right)+60 \sqrt{6}+1675}}{2 \sqrt{6} E-2 \left(24 A+\sqrt{6}+24\right)}+\frac{1}{2},
\end{align*}
\end{widetext}

where $A= \sin \left(\sqrt{3} \omega t\right)$,
$B=\sin \left(2 \sqrt{3} \omega t\right)$,
$C=\cos \left(3 \sqrt{3} \omega t\right)$,
$D=\cos \left(2 \sqrt{3} \omega t\right)$ and
$E=\cos \left(\sqrt{3} \omega t\right)$.

{\section*{Appendix B: Derivation of the QSL bound (\ref{many-body6}) for the spin-$\frac{1}{2}$ XXZ model}
Set \( L=2 \) and \( p=1/4 \). The system is initially prepared in a mixed state given by
 \begin{equation}
\rho_0^A =
\begin{pmatrix}
 \frac{5}{8} & 0 \\
 0 & \frac{3}{8}
\end{pmatrix}.
\end{equation}
{Time--evolved reduced state.}
Since $U_t\,\mathbb{I}_4\,U_t^\dagger=\mathbb{I}_4$ and $U_t\,|\Phi\rangle\!\langle\Phi|\,U_t^\dagger
=|\psi_t\rangle\!\langle\psi_t|$, the global state is
\[
\rho_t=\frac{1-p}{4}\,\mathbb{I}_4+p\,|\psi_t\rangle\!\langle\psi_t|.
\]
Taking the partial trace over $B$ gives
\[
\rho_t^A=\mathrm{tr}_B(\rho_t)=\frac{1-p}{2}\,\mathbb{I}_2
+p\,\mathrm{tr}_B\big(|\psi_t\rangle\!\langle\psi_t|\big).
\]
Hence
\[
\rho_t^A=\frac{1-p}{2}\,\mathbb{I}_2
+p\Big[\sin^2(2Jt)\,|0\rangle\!\langle0|+\cos^2(2Jt)\,|1\rangle\!\langle1|\Big],
\]
i.e.,
\[
\rho_t^A=\begin{pmatrix}
\frac{1-p}{2}+p\sin^2(2Jt) & 0\\[4pt]
0 & \frac{1-p}{2}+p\cos^2(2Jt)
\end{pmatrix}.
\]
For $p=\tfrac14$ this reduces to
\[
\rho_t^A=\mathrm{diag}\!\left(\tfrac{3}{8}+\tfrac14\sin^2(2Jt),\,\tfrac{3}{8}+\tfrac14\cos^2(2Jt)\right).
\]
These expressions can be used directly in Eq.~(\ref{many-body6}) to evaluate
$S_{q,z}(\rho_t^A)$ and $\lambda_{\min}(\rho_t^A)=\min\!\big\{\tfrac{1-p}{2}+p\sin^2(2Jt),\,
\tfrac{1-p}{2}+p\cos^2(2Jt)\big\}$, and thus the corresponding QSL bound.

\section*{Appendix C: Limiting forms  of the Sharma--Mittal entropy}
Let $h_q(\rho)=\mathrm{tr}(\rho^q)=\sum_i \lambda_i^q$, where $\{\lambda_i\}$ are
the eigenvalues of $\rho$.
\begin{itemize}
\item  $z\to1$ (R\'{e}nyi entropy)
\begin{align*}
\mathrm{S}_{q,z}(\rho)
&=\frac{1}{1-z}\!\left[\exp\!\left(\frac{1-z}{1-q}\ln h_q(\rho)\right)-1\right].
\end{align*}
Set $\varepsilon=1-z$. Using $e^{\varepsilon x}=1+\varepsilon x+o(\varepsilon)$, we have
\begin{align*}
\lim_{z\to1}\mathrm{S}_{q,z}(\rho)
=\lim_{\varepsilon\to0}\frac{e^{\varepsilon\frac{\ln h_q}{1-q}}-1}{\varepsilon}
=\frac{1}{1-q}\ln h_q(\rho),
\end{align*}
which gives Eq.~(\ref{Renyi}).
  \item $z\to q$ (Tsallis entropy)
Since $\tfrac{1-z}{1-q}\xrightarrow{}1 (z\to q)$,
\begin{align*}
\lim_{z\to q}\mathrm{S}_{q,z}(\rho)
=\frac{h_q(\rho)-1}{1-q},
\end{align*}
which yields Eq.~(\ref{Tsallis}).
  \item  $q\to1,\;z\to1$ (von Neumann entropy) $z\to1$ gives rise to the R\'{e}nyi form. Hence,
\begin{align*}
\lim_{q\to1}\frac{1}{1-q}\ln h_q(\rho)
= -\left.\frac{d}{dq}\ln h_q(\rho)\right|_{q=1}
= -\frac{h'_1(\rho)}{h_1(\rho)}.
\end{align*}
Since $h_1(\rho)=\mathrm{tr}\,\rho=1$ and $h'_q(\rho)=\mathrm{tr}\,\rho^q\ln\rho$,
\begin{align*}
\lim_{q\to1,\,z\to1}\mathrm{S}_{q,z}(\rho)=-\mathrm{tr}(\rho\ln\rho),
\end{align*}
which is Eq.~(\ref{Shannon}).
\end{itemize}

\end{document}